\documentstyle[hhline,natbib2,natbibmnfix,epsfig,lscape]{mn}
\bibpunct[,]{(}{)}{;}{a}{}{}

\newif\ifAMStwofonts

\def\tcdm{$\tau$CDM }
\def\lcdm{$\Lambda$CDM }
\def\tcdmDot{$\tau$CDM. }
\def\lcdmDot{$\Lambda$CDM. }
\def\tcdmCom{$\tau$CDM, }
\def\lcdmCom{$\Lambda$CDM, }

\def\Bband{$B$-band }

\def\BbandCom{$B$-band, }

\def\Vband{$V$-band }

\def\Iband{$I$-band }

\def\bj{$b_{j}$ }
\def\bjband{$b_{j}$-band }

\def\bjbandCom{$b_{j}$-band, }

\newcommand{\hkpc}{\mbox{$h^{-1}$ kpc} }
\newcommand{\hkpcKet}{\mbox{$h^{-1}$ kpc)} }
\newcommand{\hMpc}{\mbox{$h^{-1}$ Mpc} }
\newcommand{\hMpcCom}{\mbox{$h^{-1}$ Mpc,} }
\newcommand{\hMpcDot}{\mbox{$h^{-1}$ Mpc.} }

\newcommand{\hmsunKet}{\mbox{$h^{-1}$ $M_{\odot}$)} }
\newcommand{\hmsunCom}{\mbox{$h^{-1}$ $M_{\odot}$,} }

\newcommand{\kms}{\mbox{km s$^{-1}$} }
\newcommand{\kmsCom}{\mbox{km s$^{-1}$,} }
\newcommand{\kmsDot}{\mbox{km s$^{-1}$.} }
\newcommand{\kmsKet}{\mbox{km s$^{-1}$)} }

\newcommand{\msunyr}{\mbox{$M_{\odot}/\rmn{yr}$} }

\newcommand{\msunyrKC}{\mbox{$M_{\odot}/\rmn{yr}$),} }

\newcommand{\Mvir}{\mbox{$M_{\rmn{200}}$} }
\newcommand{\Rvir}{\mbox{$R_{\rmn{200}}$} }
\newcommand{\Vvir}{\mbox{$V_{\rmn{200}}$} }

\newcommand{\MvirCom}{\mbox{$M_{\rmn{200}}$,} }
\newcommand{\RvirCom}{\mbox{$R_{\rmn{200}}$,} }

\newcommand{\RvirKC}{\mbox{$R_{\rmn{200}}$),} }
\newcommand{\VvirD}{\mbox{$V^{2}_{\rmn{200}}$} }
\newcommand{\Vdisk}{\mbox{$V_{\rmn{disk}}$} }
\newcommand{\VdiskCom}{\mbox{$V_{\rmn{disk}}$,} }
\newcommand{\VdiskDot}{\mbox{$V_{\rmn{disk}}$.} }
\newcommand{\VdiskD}{\mbox{$V^{2}_{\rmn{disk}}$} }

\newcommand{\Lstar}{\mbox{$L_{*}$} }

\def\fbulge{f_{\rmn{bulge}}}

\def\fbar{f_{\rmn{bar}}}

\newcommand{\IRAS} {\emph{IRAS}\ }

\newcommand{\PSCZ} {\emph{PSCz}\ }
\newcommand{\UZC} {\emph{UZC}\ }

\newcommand{\APM} {\emph{APM}\ }
\newcommand{\GIF} {\emph{GIF}\ }

\newcommand{\MarkIII} {\emph{Mark III}\ }
\newcommand{\SFI} {\emph{SFI}\ }

\newcommand{\PSCZCom} {\emph{PSCz,}\ }
\newcommand{\UZCCom} {\emph{UZC,}\ }

\newcommand{\MarkIIIDot} {\emph{Mark III.}\ }

\newcommand{\gadget} {{\sc gadget}\ }

\newcommand{\POTENT} {{\sc POTENT}\ }

\def\la{\mathrel{\hbox{\rlap{\hbox{\lower4pt\hbox{$\sim$}}}\hbox{$<$}}}}
\def\ga{\mathrel{\hbox{\rlap{\hbox{\lower4pt\hbox{$\sim$}}}\hbox{$>$}}}}

\def\lsim{\la}

\newcommand{\bspa}{\begin{spacing}{2}}
\newcommand{\espa}{\end{spacing}}

\ifoldfss
  \newcommand{\rmn}[1] {{\rm #1}}

  \ifCUPmtlplainloaded \else
    \NewTextAlphabet{textbfit} {cmbxti10} {}
    \NewTextAlphabet{textbfss} {cmssbx10} {}
    \NewMathAlphabet{mathbfit} {cmbxti10} {} % for math mode
    \NewMathAlphabet{mathbfss} {cmssbx10} {} %  "   "    "
  \fi
  \ifAMStwofonts
    \ifCUPmtlplainloaded \else
      \NewSymbolFont{upmath} {eurm10}
      \NewSymbolFont{AMSa} {msam10}
      \NewMathSymbol{\upi}     {0}{upmath}{19}
      \NewMathSymbol{\umu}     {0}{upmath}{16}
      \NewMathSymbol{\upartial}{0}{upmath}{40}
      \NewMathSymbol{\leqslant}{3}{AMSa}{36}
      \NewMathSymbol{\geqslant}{3}{AMSa}{3E}

    \fi
  \fi
\fi % End of OFSS

\ifnfssone
  \newmathalphabet{\mathit}
  \addtoversion{normal}{\mathit}{cmr}{m}{it}
  \addtoversion{bold}{\mathit}{cmr}{bx}{it}
  \newcommand{\rmn}[1] {\mathrm{#1}}

  \newmathalphabet{\mathbfit} % math mode version of \textbfit{..}
  \addtoversion{normal}{\mathbfit}{cmr}{bx}{it}
  \addtoversion{bold}{\mathbfit}{cmr}{bx}{it}
  \newmathalphabet{\mathbfss} % math mode version of \textbfss{..}
  \addtoversion{normal}{\mathbfss}{cmss}{bx}{n}
  \addtoversion{bold}{\mathbfss}{cmss}{bx}{n}
  \ifAMStwofonts
    \ifCUPmtlplainloaded \else
      \UseAMStwoboldmath
      \makeatletter
      \new@mathgroup\upmath@group
      \define@mathgroup\mv@normal\upmath@group{eur}{m}{n}
      \define@mathgroup\mv@bold\upmath@group{eur}{b}{n}
      \edef\UPM{\hexnumber\upmath@group}
      \new@mathgroup\amsa@group
      \define@mathgroup\mv@normal\amsa@group{msa}{m}{n}
      \define@mathgroup\mv@bold\amsa@group{msa}{m}{n}
      \edef\AMSa{\hexnumber\amsa@group}
      \makeatother
      \mathchardef\upi="0\UPM19
      \mathchardef\umu="0\UPM16
      \mathchardef\upartial="0\UPM40
      \mathchardef\leqslant="3\AMSa36
      \mathchardef\geqslant="3\AMSa3E
    \fi
  \fi
\fi % End of NFSS release 1

\ifnfsstwo
  \newcommand{\rmn}[1] {\mathrm{#1}}

  \DeclareMathAlphabet{\mathbfit}{OT1}{cmr}{bx}{it}
  \SetMathAlphabet\mathbfit{bold}{OT1}{cmr}{bx}{it}
  \DeclareMathAlphabet{\mathbfss}{OT1}{cmss}{bx}{n}
  \SetMathAlphabet\mathbfss{bold}{OT1}{cmss}{bx}{n}
  \ifAMStwofonts
    \ifCUPmtlplainloaded \else
      \DeclareSymbolFont{UPM}{U}{eur}{m}{n}
      \SetSymbolFont{UPM}{bold}{U}{eur}{b}{n}
      \DeclareSymbolFont{AMSa}{U}{msa}{m}{n}
      \DeclareMathSymbol{\upi}{0}{UPM}{"19}
      \DeclareMathSymbol{\umu}{0}{UPM}{"16}
      \DeclareMathSymbol{\upartial}{0}{UPM}{"40}
      \DeclareMathSymbol{\leqslant}{3}{AMSa}{"36}
      \DeclareMathSymbol{\geqslant}{3}{AMSa}{"3E}
    \fi
  \fi
\fi % End of NFSS release 2

\ifCUPmtlplainloaded \else
  \ifAMStwofonts \else % If no AMS fonts
    \def\upi{\pi}
    \def\umu{\mu}
    \def\upartial{\partial}
  \fi
\fi

\title{Simulating the Formation of the Local Galaxy Population}
\author[H. Mathis et al.]
	{H.~Mathis,$^1$\thanks{Email: hmathis@mpa-garching.mpg.de} 
	G.~Lemson,$^2$
	V.~Springel,$^1$  
	G.~Kauffmann,$^1$
	S.~D.~M.~White,$^1$  	
	\newauthor A.~Eldar~$^2$ and A.~Dekel~$^2$
	\\	
        $^1$Max--Planck--Institut f\"ur Astrophysik, D-85741 Garching, Germany
	\\
        $^2$Racah Institute of Physics, The Hebrew University, Jerusalem, Israel}
	
\pagerange{\pageref{firstpage}--\pageref{lastpage}}

\begin{document}

\maketitle

\label{firstpage}

%%%%%%%%%%%%%%%%%%%%%%%%%%%%%%%%%%%%%%%%%%%%%%%%%%%%%%%%%%%%%%%%%%%%%%%%%%%%%%%%%%%%%%%%%%%%%

\begin{abstract}

We simulate the formation and evolution of the local galaxy
population starting from initial conditions with a smoothed linear 
density field which matches that derived from the \IRAS 1.2 Jy galaxy 
survey. Our simulations track the formation and evolution of all dark 
matter haloes more massive than $10^{11}\rmn{M}_\odot$ out to a distance 
of 8000 \kms from the Milky Way. We implement prescriptions similar 
to those of \citet{Kau99} to follow the assembly and evolution of 
the galaxies within these haloes. We focus on two variants of the 
CDM cosmology: a \lcdm and a \tcdm model. Galaxy formation in each
is adjusted to reproduce the \Iband Tully--Fisher relation of 
\citet{Gio97}. We compare the present-day luminosity functions, 
colours, morphology and spatial distribution of our simulated 
galaxies with those of the real local population, in particular 
with the Updated Zwicky Catalog, with the \IRAS  \PSCZ redshift 
survey, and with individual local clusters such as Coma, Virgo and 
Perseus. We also use the simulations to study the clustering bias 
between the dark matter and galaxies of differing type. Although
some significant discrepancies remain, our simulations recover the
observed intrinsic properties and the observed spatial distribution 
of local galaxies reasonably well. They can thus be used to calibrate
methods which use the observed local galaxy population to estimate
the cosmic density parameter or to draw conclusions about the
mechanisms of galaxy formation. To facilitate such work, we
publically release our $z=0$ galaxy catalogues, together with
the underlying mass distribution.

\end{abstract}
 

\section[]{Introduction}
\label{sec:Intro}

Over the last decade phenomenological modelling has made
it possible to follow many aspects of the formation and evolution of 
galaxies within the currently favored hierarchical paradigm for the 
growth of cosmic structure. Recently, the grafting of techniques
originally developed by \citet{Wh91,Kau93b,Col94} onto high resolution N-body simulations has
allowed the spatial and kinematic distributions of galaxies to be
predicted in detail as a function of their intrinsic properties 
\citep{Kau97,Kau99,Kau99b,Dia99,Dia01,Ben00a,Ben00b,Ben01a,Ben01b,Spr00}.
This work clarifies many
aspects of the problem of `galaxy biasing' and supersedes the 
heuristic models previously used to relate the galaxy and mass 
distributions in CDM cosmogonies. Its goal is twofold: to better 
understand the physical processes driving the formation, evolution 
and clustering of galaxies, and to test the standard structure
formation paradigm.

With the steady improvement of computer performance and of simulation
codes, dissipationless simulations are able to achieve ever higher 
mass resolution. In a recent example, \citet[hereafter S00]{Spr00} followed
the formation of all galaxies brighter than the Fornax 
dwarf spheroidal within a cluster similar in mass to Coma. 
In the simulations presented below, the use of $7\times 10^7$
particles allows us to follow the formation of all galaxies more 
luminous than the LMC out to 8000 \kms from the Milky Way. Improved
computers have also greatly enhanced the ability of cosmic 
gas-dynamics codes to simulate galaxy formation (e.g. \citealt{Pear01,Nag00,WhiM00}). 
Although such simulations remove the uncertainties due to phenomenological
modelling of the dynamics and cooling of diffuse gas,
they retain phenomenological models for the much more uncertain 
processes of star formation and supernova feedback. Moreover, their 
greatly increased computational cost makes it impossible for them 
to resolve galaxy formation over a volume as large as that modelled 
in this paper. A further major advantage of the techniques used here is
that the efficiencies for uncertain processes
like star formation and feedback can be varied to study their
influence and then adjusted to fit observation without the need 
to run a new simulation for each new parameter set.
 
The simulations we present below use initial conditions generated
using a technique developed by \citet[see \citet{Bi98}
for a related technique]{Kol96}. An all-sky
redshift survey is smoothed heavily to produce an estimate of the
galaxy overdensity field in a spherical volume centred on the Milky Way.
This is assumed to be a known constant times the similarly smoothed 
local mass overdensity field. One then solves for the linear 
overdensity field at high redshift which would evolve into this
quasi-linear local field. The initial conditions for the 
simulation are taken to be a random realisation of a Gaussian
random field with a suitable CDM (or other) power spectrum, but
{\it constrained} so that when suitably smoothed the overdensity 
field is equal to that inferred from the local
galaxy distribution. Simulations run from such initial conditions
reproduce the large-scale structure of the local universe but have
the characteristics of the assumed CDM model on small scales where
the smoothed local galaxy distribution imposes no constraints.

Although galaxy formation should occur in such simulations exactly
as in random realisations of the underlying CDM model, there are a 
number of advantages to having the large-scale structure of the
simulation correspond in detail to that of the local universe. For certain types
of galaxies, in particular dwarfs, surveys are restricted to our
local neighborhood. For many types of galaxies, surveys are most 
complete and the properties of the galaxies best characterised
in this region. When interpreting such surveys one must be wary of 
biases introduced by the particular structure of our neighborhood. 
Such biases are clearly minimised in models which reproduce the 
local structure. Distances to galaxies can only be measured with 
sufficient accuracy to estimate their peculiar motions out to
redshifts of about 10,000 \kmsDot As a result, detailed studies of 
large-scale flows are only possible in our local neighborhood.
Such studies aim to verify that flows are gravitationally induced 
and to use them to measure the cosmic density parameter. Simulations
of the nearby universe are ideal for calibrating such studies and
for checking that they produce unbiased estimates of $\Omega_m$.

The present paper is organized as follows. In Section 2 we describe
both how we construct constrained initial conditions from the 
density field of the \IRAS 1.2 Jy survey and how we carry out dark
matter simulations from these initial conditions. Section 3
explains how halo catalogues and halo merging trees are built
from the simulation outputs and summarises the phenomenological
treatment of galaxy formation which we graft onto these trees,
emphasising points where it differs from the treatment in
\citet[hereafter K99]{Kau99}. This section also compares our dark matter distributions
to those found in unconstrained simulations and our simulated
distributions of galaxy luminosity, colour and morphology to those
observed in large surveys. Section 4 begins our detailed 
comparison with the local universe by matching simulated rich 
clusters object by object with real rich clusters. In Section
5 we explain how we generate `mock' catalogues for direct 
comparison with the \IRAS \PSCZ and \UZC surveys. In Section 6 
we use these catalogues to carry out a point-by-point comparison 
of the smoothed galaxy and mass density fields, while Section 7 
extends the comparison to smaller scales using cross-correlation 
statistics. Section 8 displays mock versions of the \MarkIII
catalogue of galaxy peculiar velocities to illustrate the application
of our simulations to cosmic flow problems. Finally, we give a brief 
summary and discussion of our results in Section 9.

%%%%%%%%%%%%%%%%%%%%%%%%%%%%%%%%%%%%%%%%%%%%%%%%%%%%%%%%%%%%%%%%%%%%%%%%%%%%%%%%%%%%%%%%%%%%%

\section[]{Dark Matter}
\label{sec:DM}

%%%%%%%%%%%%%%%%%%%%%%%%%%%%%%%%%%%%%%%%%%%%%%%%%%%%%%%%%%%%%%%%%%%%%%%%%%%%%%%%%%%%%%%%%%%%%

Our modelling is based on large high resolution simulations of the
evolution of the dark matter distribution in a spherical region
surrounding the Milky Way. The first half of this section 
explains how we obtain initial conditions for these simulations;
the second half describes the code used to follow their evolution.

\subsection{Constructing initial conditions}

We use the techniques developed by \citet[hereafter K96]{Kol96} to set
up initial conditions such that at $z=0$ the evolved mass 
overdensity in the simulation is a suitably
scaled version of the galaxy overdensity in the \IRAS 
1.2 Jy survey of \citet{Fi94,Fi95} once both are smoothed on
the same large scale.  We take all the observed galaxies 
out to a redshift of 12,000 \kmsCom weight each by the
inverse of the survey selection function at its redshift, 
and smooth with a Gaussian of 1-D dispersion $5\:\hMpc$ to
obtain the galaxy density field at all points
within a sphere of radius 10,000 \kmsDot When smoothing 
we take care to account properly for the regions of  
space not sampled by the observational survey. These are
primarily behind the Galactic Plane and in the strip
which was not scanned by the \IRAS satellite. 
	
For each cosmology (\lcdm and \tcdm as in K99) 
we obtain a `target' mass overdensity 
field by scaling the galaxy overdensities so that their 
{\it rms} within the region out to 10,000 \kms is equal 
to the value obtained by smoothing the $z=0$ linear power 
spectrum of the particular cosmological model with a Gaussian
of dispersion $5\:\hMpcDot$ As described by K96 the
Eulerian Zel'dovich-Bernoulli equation can then be integrated
back in time to give the linear density field at $z=50$
which gives rise to the target smooth density field at the
present day. This linear field is then ``Gaussianised'', 
i.e. it is mapped onto a new field such that the new density at each
point is a monotonic function of the original overdensity and
is Gaussian on the observed region with {\it rms} equal to that 
expected at $z=50$ from the smoothed theoretical power spectrum.

We now apply the Hoffman-Riback algorithm \citep{Hof91,Hof92,Gan93}
to generate an initial displacement field on a 256$^3$ grid
which (i) is periodic on a simulation cube of side $L=240$ \hMpcCom (ii) is a Gaussian random 
field with power per mode equal to that expected at $z=50$
for the chosen cosmology for all wavenumbers between the
fundamental, $k_{0} = 2\pi/L$ and $64\:k_{0}$, and (iii) is constrained
so that when smoothed with a Gaussian of comoving dispersion
$5\:\hMpc$ it reproduces the target initial density field everywhere
within a sphere of radius 80 \hMpc centred on the cube.

We supplement this constrained low-frequency displacement field
with an unconstrained high-frequency displacement field constructed
as follows. On a periodic cube of side 80 \hMpc we set up an
unconstrained Gaussian random field with power per mode equal
to that expected in the chosen cosmology for all wavenumbers
between $64\:k_{0}$ and \mbox{$2\pi/(0.7 \hMpc)= 343\:k_{0}$}. 
The latter corresponds to a wavelength equal to twice the
mean interparticle separation of our high resolution region
(see below). We replicate this smaller cube 27 times to
obtain a high-frequency displacement field everywhere within 
the 240 \hMpc cube.

Because the version of the integration code we use has vacuum boundary
conditions, it works best on near-spherical regions. We therefore
create an unperturbed but variable resolution particle load within a
sphere of radius  $ 240\:\hMpc \times \sqrt{3}/2 =  207.85 \:\hMpcDot$ We
first fill the 240 \hMpc cube with a uniform but irregular ``glass''
of equal mass particles. (See White (1995) for a discussion of such glass
distributions.) We then border this cube with periodic replications
and select the sphere which just encloses the central cube as the
final region to be simulated. Within this large sphere, we excise all
particles which will end up less than 80 \hMpc from the center by
$z=0$, and we replace them with a region of the same size and shape
cut from a glass distribution with a mean interparticle separation of 
0.35 \hMpcDot The particle masses are adjusted to ensure the correct
mean mass density in each of the two regions. Table~\ref{tab:SIMPar} 
lists the cosmological parameters and the particle numbers 
and masses for each of our two simulations.

The final initial condition (at $z=50$) is created by interpolating the
high- and low-frequency displacement fields to the unperturbed
position of each particle, displacing it, and assigning it a 
peculiar velocity proportional to its displacement in accordance with 
usual Zel'dovich approximation. We move the low-mass, high-resolution
particles in the inner region using the sum of the two displacement
fields. The high-mass, low-resolution particles in the
outer region are perturbed using the low-frequency constrained 
displacement field only. The latter field must be periodically
replicated in order to displace the low resolution particles outside
the 240 \hMpc central cube.

%%%%%%%%%%%%%%%%%%%%%%%%%%%%%%%%%%%%%%%%%%%%%%%%%%%%%%%%%%%%%%%%%%%%%%%%%%%%%%%%%%%%%%%%%%%%%

%%%%%%%%%%%%%%%%%%%%%%%%%%%%%%%%%%%%%%%%%%%%%%%%%%%%%%%%%%%%%%%%%%%%%%%%%%%%%%%%%%%%%%%%%%%%%

\subsection{Simulating the Dark Matter}

\begin{table*}
\begin{minipage}{160mm}
\centering
\caption{Simulation parameters of our dissipationless cosmological
simulations. $\rmn{N_{hr}}$, $\rmn{N_{lr}}$, $\rmn{M_{hr}}$,
$\rmn{M_{lr}}$ and $r_{\rmn{soft,hr}}$, $r_{\rmn{soft,lr}}$ give the number, 
the mass (units \hmsunKet and the {\it physical} softening lengh (units \hkpcKet of the high-resolution and 
low-resolution particles, respectively. }
\begin{tabular}{@{}lllllllllll@{}}
\hline
Model & $\rmn{h}_{\rmn{100}}$ & $\Omega_{\rmn{0}}$ & $\Lambda$ & $\sigma_{8}$ &
$\rmn{N_{hr}}$ & $\rmn{M_{hr}} \mskip+12mu $   &
$r_{\rmn{soft,hr}}$ & 
$\rmn{N_{lr}}$ & $\rmn{M_{lr}} \mskip+12mu $   & 
$r_{\rmn{soft,lr}}$  \\
\hline
\rule{0in}{3ex}
$\Lambda$CDM & 0.7 & 0.3 & 0.7 & 0.9 & $51 \times 10^{6}$ & $0.36
\times 10^{10}$ & 20 & $20.5 \times 10^{6}$ & $1.4 \times 10^{11}$ & 120 \\
\hline
\rule{0in}{3ex}
$\tau$CDM & 0.5 & 1.0 & 0.0 & 0.6 & $53 \times 10^{6}$ & $1.2
\times 10^{10}$ & 20 & $20.4 \times 10^{6}$ & $ 4.8 \times 10^{11}$ & 120 \\
\hline
\end{tabular}
\label{tab:SIMPar}
\end{minipage}
\end{table*}
  
We carried out simulations from the above initial conditions 
using the parallel tree-code \gadget \citep{SprGadget2001}. 
Both simulations were started with fixed comoving softenings of 70 
\hkpc and 420 \hkpc for high- and low-resolution particles 
respectively. Once the corresponding physical comoving softenings
reached 20 \hkpc and 120 \hkpc they were kept constant at these values. 

For the \lcdm simulation, this resulted 
in particles in the densest regions of the high resolution region 
requiring roughly 75000 adaptive timesteps to reach redshift
zero. The number of equivalent full timesteps (the total number of
forces computed divided by the number of particles) was much smaller however,
of order 400. Both runs were performed on 512 processors of the CRAY T3E at 
the Computer Center (RZG) of the Max-Planck Society at Garching.   
The \lcdm simulation required about 104 processor-hours to complete
(the total CPU time consumption was about 50000 hours), while the \tcdm
simulation was somewhat faster. We saved the 
particle positions and velocities at 41 times 
logarithmically spaced in expansion parameter from $z=10.3$ to $z=0$.

%%%%%%%%%%%%%%%%%%%%%%%%%%%%%%%%%%%%%%%%%%%%%%%%%%%%%%%%%%%%%%%%%%%%%%%%%%%%%%%%%%%%%%%%%%%%%

% DMFIG
	
\begin{figure*}
\begin{minipage}{180mm}
\centering
\caption{The present-day distribution of the dark matter. \lcdm and 
\tcdm are shown on left and right, respectively.  
The slice shown is $180\: \hMpc$ wide, $30\: \hMpc$ thick and encompasses
the supergalactic plane. Prominent clusters are labelled. A high resolution copy of this Figure
can be found at http://www.mpa-garching.mpg.de/NumCos/CR/High-res/index.html}
\label{fig:DMDistFig}
\end{minipage}
\end{figure*}

%%%%%%%%%%%%%%%%%%%%%%%%%%%%%%%%%%%%%%%%%%%%%%%%%%%%%%%%%%%%%%%%%%%%%%%%%%%%%%%%%%%%%%%%%%%%%

%%%%%%%%%%%%%%%%%%%%%%%%%%%%%%%%%%%%%%%%%%%%%%%%%%%%%%%%%%%%%%%%%%%%%%%%%%%%%%%%%%%%%%%%%%%%%

In Fig.~\ref{fig:DMDistFig} we show the
present-day dark matter distribution in our \lcdm and \tcdm models.
The projected region is a 30 \hMpc thick slice parallel to the
supergalactic plane and extending from -15 \hMpc to 15 \hMpc in SGZ. The
side of the slice is 180 \hMpc long, and the Milky-Way is at its
centre. Well-known local structures are easily identified, in
particular, the Great Attractor region, the Coma-Great Wall structure,
the Pisces-Perseus supercluster filament, and the Local Void. 
 
Many of the known nearby rich clusters, including
Virgo, can also be identified, even though the smoothing kernel
used to generate the initial conditions is substantially larger than
a rich cluster and such clusters 
are actually quite poorly represented in the \IRAS 1.2 Jy survey.

One can start to see the low resolution
region at the corners of these images which extend
more than 80 \hMpc from the centre. In the remainder of this paper
we shall concentrate on analysis of the region within 80 \hMpc and
we will exclude from consideration the few dark haloes near the
boundary which are contaminated with high-mass particles.

\begin{figure}					
	\centering
	\epsfig{file=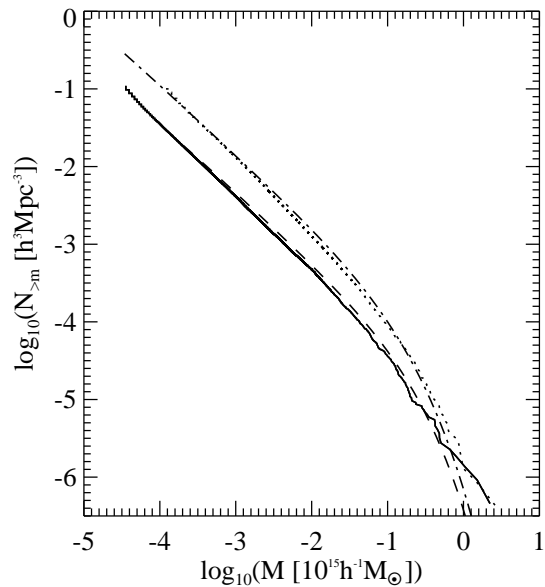,width=8.5cm,height=8.5cm}						
	\caption {Mass function of haloes in the simulations. The
	solid and dotted lines are the simulated \lcdm and \tcdm
	models respectively. The dashed and dashed-dotted lines are
	the \lcdm and \tcdm predictions of \citet{Je00}.}
	\label{fig:MFFig}
\end{figure}

As a test that our simulations are correctly reproducing the small
scale structure expected in each cosmology, Fig.~\ref{fig:MFFig}
compares their halo mass functions at $z=0$ with those expected
according to the large ensemble of unconstrained simulations
analysed by \citet{Je00}. As discussed in the next section, we identify
haloes in our simulations using a friends-of-friends group-finder
with $b=0.2$ \citep{Da85}. The analytic fit which \citet{Je00} use
to summarise their data for such FOF mass functions is also an
excellent description of our own halo data. Note that no parameters 
were adjusted in making this comparison.  Significant
discrepancies occur only for the largest mass haloes. These are
plausibly a consequence of the constraints we impose on our initial
conditions.

%%%%%%%%%%%%%%%%%%%%%%%%%%%%%%%%%%%%%%%%%%%%%%%%%%%%%%%%%%%%%%%%%%%%%%%%%%%%%%%%%%%%%%%%%%%%%

%%%%%%%%%%%%%%%%%%%%%%%%%%%%%%%%%%%%%%%%%%%%%%%%%%%%%%%%%%%%%%%%%%%%%%%%%%%%%%%%%%%%%%%%%%%%%

\section[]{Putting in the Galaxies}
\label{sec:SA}

We follow the assembly, the evolution and the merging of galaxies
within our dark matter simulations using phenomenological
prescriptions based closely on those of K99. 
In this section we give an outline description of the
scheme as a whole. We then discuss particular aspects of the
modelling where we have found it advantageous to modify the K99 
prescriptions. Finally we describe how we set the parameters which
describe uncertain physical processes (for example, the efficiencies
of star formation and feedback) using observed properties of galaxies
such as the Tully--Fisher relation and the luminosity function.

Our galaxy formation scheme is based on the idea, originally
introduced by \citet{Whi78}, that galaxies result from the
condensation of diffuse gas at the centres of dark haloes. As a
result all galaxies are found within dark haloes, either as the
principal central object or as satellites in orbit about
it. The first step is thus to identify all the dark haloes in each
snapshot of the simulation and to construct a tree which 
describes how they merge together from snapshot to snapshot. 
The galaxy population can be reconstructed by following evolution
down the tree from the earliest times to the present. Within 
each halo at each time diffuse gas cools onto the disk of the central 
galaxy, stars form from cold gas both in this galaxy and in its 
satellites, feedback from this star formation reheats some of the 
cold gas, and satellites occasionally merge with the central galaxy,
modifying its morphology if they are sufficiently massive. Galaxy 
motions are followed in a dynamically consistent way by attaching 
each galaxy to a dark matter particle. For a central galaxy 
this identification is with the most bound particle of the halo.
The particle identification remains fixed if this halo merges with a more 
massive one so that the galaxy becomes a satellite.

We now briefly discuss aspects of this modelling, emphasising aspects 
where our implementation differs in detail from that of K99.

%%%%%%%%%%%%%%%%%%%%%%%%%%%%%%%%%%%%%%%%%%%%%%%%%%%%%%%%%%%%%%%%%%%%%%%%%%%%%%%%%%%%%%%%%%%%%

\subsection{Tree structure}
\label{sec:SA:Tree}

As in K99 we build a halo catalogue for each of the 41 
snapshots of each simulation  using a friends-of-friends algorithm 
with linking length 0.2 times the mean interparticle separation 
\citep{Da85}. For further analysis we consider only haloes with 10 
or more particles; smaller haloes frequently disappear in later
snapshots. The more massive haloes are linked from each snapshot 
to the next to form a tree exactly as in K99. Note that although 
we can pinpoint the location and mass of haloes quite well down 
to our limit of 10 particles, we can only reliably track the 
merging history of haloes more massive than about 100 particles.
Lower mass haloes can typically be traced back through a few 
snapshots only before they vanish below the resolution limit. 
As a result, we can get reasonable estimates for the mass of 
the central galaxy in a 10 to 20 particle halo, but we cannot 
infer anything about its morphology for halo masses below 100.

This leads us to define the \emph{luminosity} resolution limit of
our simulated galaxy catalogues to be the typical luminosity of the
central galaxy in a 10 particle halo, and the \emph{morphology} 
resolution limit to be the typical luminosity of the central galaxy in
a 100 particle halo. At $z=0$ the circular velocities of 10 and
100 particle haloes are 45 \kms and 103 \kms respectively in our 
\lcdm simulation. For \tcdm the corresponding numbers are 76 \kms 
and 160 \kmsDot (We define the circular velocity \Vvir of a halo to be
the value measured at \RvirCom the radius of the sphere centred on the
most bound particle within which the mean density is 200 times the
{\it critical} density of the Universe.) In the galaxy catalogues  
 we construct below, the corresponding \emph{luminosity} and 
\emph{morphology} limits are $M_{B}=-16.27$ and $-18.46$
for \lcdmCom $M_{B}=-18.50$ and $-20.60$ for \tcdmDot
In this paper, ``resolution limit'' without a qualifier 
will always denote the \emph{luminosity} resolution limit.

Galaxies are of three types in our evolution scheme: {\it central} 
galaxies which are always identified with the most bound particle 
(MBP) of their halo, {\it satellite} galaxies which are identified with 
other halo particles, and {\it field} galaxies whose particle is 
not a member of any catalogued halo. When setting up the galaxy 
population at redshift $z_0$, we adopt the following 
procedures which differ slightly from those of K99.
	\begin{enumerate}
	\renewcommand{\theenumi}{(\arabic{enumi})}
	\item If no particle in a halo was a galaxy in the
previous snapshot at $z_1>z_0$, then we create a new
galaxy with default properties based on a simple infall model 
for the earlier evolution of the halo up to $z_0$.
	\item If no particle in the halo was the central galaxy
of a halo at $z_1$, we identify the most massive (in terms 
of total stellar and cold gas mass) of the galaxies which do
populate the halo as its central galaxy, we transfer the 
properties of this galaxy to the halo MBP, and we delete the 
particle previously assigned to it from the galaxy list.
        \item If at least one particle was a central galaxy at $z_1$,
we identify the new central galaxy as the one whose halo donated the 
most mass to the $z_0$ halo. Its properties are transferred to the MBP
and its previous particle identification is deleted from the galaxy
list. The other `old' central galaxies become satellite galaxies in
the new halo.
        \item Galaxies from $z_1$ which are part of no halo at
$z_0$ are designated as field galaxies. These objects are always
few in number and are near the resolution limit of the simulation.
They play no significant role in the results we present below.
        
	\end{enumerate}	

This scheme is designed to ensure that a halo central galaxy always
inherits the properties of the most plausible (usually the most
massive) candidate for its progenitor in the previous snapshot.
We have tested a variant in which inheritance is based on luminosity
rather than mass and find it to make no discernable difference. 
The scheme corrects a loophole in the K99 procedures which allowed
relatively massive haloes occasionally to find no progenitor for 
their central galaxies. As a result they were assigned relatively
low-mass central galaxies and they appeared as outliers in the
Tully--Fisher relation.

%%%%%%%%%%%%%%%%%%%%%%%%%%%%%%%%%%%%%%%%%%%%%%%%%%%%%%%%%%%%%%%%%%%%%%%%%%%%%%%%%%%%%%%%%%%%%
		
\subsection{Cooling, star formation and feedback}

The 41 snapshots stored from $z=10.42$ to $z=0$ sample the evolution
of the system more coarsely than is desirable when simulating the
evolution of the galaxy population. As in K99 we circumvent this by 
dividing the time between each pair of snapshots into 50 small steps, 
which we use for integrating the simple phenomenological equations
which describe gas infall and cooling, star formation, feedback
and satellite merging. We have checked that doubling the number of 
steps does not change our results.

We treat cooling using exactly the same prescriptions as K99. These
assume that gas cools out of the hot atmosphere of a halo onto the 
disk at its centre at a rate which depends only on redshift, on halo 
mass and on halo hot gas content. For haloes in which
cooling is rapid (at high redshift, relatively small mass, and large
hot gas fraction) the atmosphere is assumed to condense out on the 
dynamical time of the halo. When cooling times are longer, the condensation
rate is taken from a simple cooling flow model. We calculate cooling
times using the collisional ionisation cooling curve of \citet{Su93} 
assuming primordial abundances. We do not attempt
to follow chemical enrichment. We treat the global baryon fraction
of the models $f_{\rmn{bar}}\equiv \Omega_{\rmn{bar}}/\Omega_0$ as an adjustable
parameter.

These prescriptions give rise to a problem that has been noticed at
least since the work of \citet{Wh91}. In galaxy cluster haloes
they predict strong cooling flows (similar to those apparently
{\it observed} in clusters like Perseus) which deposit large amounts 
of cold gas at cluster centres. The same star formation prescriptions 
which produce reasonable galaxies at the centre of $10^{12}\rmn{M}_\odot$
haloes then predict massive, luminous, star-forming galaxies which are
quite inconsistent with the observed properties of the central
galaxies in clusters. This is just another facet of the well-known 
problem of the fate of the gas apparently deposited by cooling
flows \citep{Fa91,All97}.

A number of authors have tried to deal with this problem in the
present context by building more elaborate models for the structure
and thermodynamic history of the gas in dark matter haloes 
\citep[hereafter SP99, C00]{Som98,Co00}. We have tried implementing 
the prescriptions advocated by SP99 and find that in 
our own models they produce only a rather limited improvement. As a
result, we have preferred to retain the simpler but cruder {\it ad
hoc} solution of K99 -- we switch off all cooling of gas
in haloes with circular velocities exceeding 350 \kmsDot This is
equivalent to assuming that gas processed through cluster-like cooling
flows does not end up being available for normal star formation. It
also ensures that no disks are formed with circular velocities
exceeding the largest observed for real disk galaxies.

Our models for star formation and feedback are also identical to
those of K99. The star formation rate in a galaxy is assumed to
be proportional to its cold gas mass divided by its dynamical time:
        \begin{equation}
        \label{eqn:sfr1}
        \dot{M}_{*}=\alpha{{M_{\rmn{cold}}}\over{t_{\rmn{dyn}}}}=100\:\alpha \: H(z) \: M_{\rmn{cold}}
	\end{equation}
where $\alpha$ is a free parameter to be fitted against observation,
$H(z)$ is the Hubble parameter, and the second equality follows from
a very simple model for the typical dynamical time of galaxy disks, 
taken to be $0.1\: \Rvir / \Vvir$. We use the population synthesis 
models of \citet{Bru93} to compute 
the luminosities of our galaxies from their star formation
histories assuming a Scalo IMF.

Feedback is assumed to act by reheating some of the cold gas in the
disk. We write the reheating rate as:
        \begin{equation}
        \label{eqn:reheatm}
        \dot{M}_{\rmn{reheat}}=\epsilon{{4}\over{3}}{{\dot{M}_{*}\eta_{\rmn{SN}}\rmn{E}_{\rmn{SN}}}\over
{\rmn{V}^{2}_{\rm{200}}}}
        \end{equation}
where $\eta_{\rmn{SN}}=5\times10^{-3} \rmn{M}^{-1}_{\odot}$ is the
number of supernovae produced per solar mass of stars in a
Scalo IMF, $\rmn{E}_{\rmn{SN}}\sim 10^{51} \rmn{erg}$ is the mean
energy released per supernova, $\rmn{V}_{\rm{200}}$ is the
circular velocity of the halo at \RvirCom and $\epsilon$ is a free 
parameter to be tuned to match observations. Since it is unclear 
whether the reheated gas will remain in the halo or will be expelled,
a model of each type was discussed by K99. We have tried both schemes
in our own simulations and, as in K99, we find that \lcdm fits the
observations better with retention feedback. For \tcdm, however,
we are able to fit the Tully--Fisher relation with less scatter using
the ejection scheme. This may appear surprising in view of fig.~6 of
K99, but it is accounted for by our different prescription for
computing \Vdisk (see below).

\subsection{Dust}

We use the simple model of dust extinction proposed by SP99. 
We compute the face--on \Bband optical depth as:
        \begin{equation}
        \label{eqn:tauB}
        \tau_{B}=\tau_{B,*}\left({{L_{B}}\over{L_{B,*}}}\right)^{\beta}
        \end{equation}
with $\tau_{B,*}=0.8$, $L_{B,*}=6 \times 10^{9}\rmn{L}_{\odot}$ and $\beta=0.5$,
as given by \citet{Wan96}. We then use the galactic extinction 
curve of \citet{Car89} to calculate the optical
depth in other bands. We pick a random inclination for each galaxy,
and apply a dust correction to its disk component only. This is calculated
using a ``slab'' model as: 
        \begin{equation}
        \label{eqn:ALambda}
        A_{\rmn{\lambda}}=-2.5 \:
\rmn{log_{10}}\left({{1-\rmn{e}^{-\tau_{\rmn{\lambda}}\sec\theta}}
        \over{\tau_{\rmn{\lambda}}\sec \theta}}\right)
        \end{equation}
K99 apply a dust correction only to the disks of galaxies with star 
formation rates $\rmn{SFR} > 0.5 \msunyr$. Here, we correct all disks,
regardless of their SFR. In practice, the differences induced by the
change in prescription are not large, and although the new prescription
gives somewhat better results, it is still unable to eliminate the 
apparent excess of galaxies at the bright end of the luminosity
function. 

\subsection{Merging}
\label{sec:SA:Rec:Merging}

As in K99, we allow a satellite galaxy to merge only with the central
galaxies of its halo. Each satellite galaxy carries a merging 
time counter which is initialized with an estimate of the timescale  
for orbital decay through dynamical friction. This
counter is reinitialized each time the satellite's halo undergoes
a major merger or is accreted by a more massive halo. Otherwise it
decreases steadily and the satellite is assumed to merge with the
central galaxy when it reaches zero. 

The initial timescale is based both on analytic estimates
\citep{Bi87} and on fits to numerical simulations \citep{Na95}.
Some problems with these formulae have been pointed out by
S00, who note that when haloes of comparable mass
merge the inferred merging times for their central galaxies
are typically much shorter than the time actually taken in
high resolution simulations of the process. This could, in principle,
lead to an overmerging problem. There is also disagreement in the 
literature on the appropriate formula for the Coulomb logarithm 
when estimating dynamical friction timescales. \citet{Som98} take
$\ln{\Lambda}=\ln(1+({M_{\rmn{halo}}}/{M_{\rmn{sat}}})^2)$ while
K99 and S00 use $\ln{\Lambda}=\ln({M_{\rmn{halo}}/M_{\rmn{sat}}})$ 
and $\ln{\Lambda}=\ln(1+{M_{\rmn{halo}}/M_{\rmn{sat}}})$
respectively. Here, we adopt the formula of SP99 
since it improves the bright--end behaviour of our luminosity functions.

We attempt to avoid any overmerging problem by 
ensuring that galaxies never merge on timescales shorter than their
halo crossing times. If $t_{\rmn{fric}}$ is the merging 
timescale derived from dynamical friction considerations, we
initialise the merging time counter using:
	\begin{equation}
	\label{eqn:merging}	
	t_{\rmn{merging}}=t_{\rmn{dyn}}(2+\frac{t_{\rmn{fric}}}{t_{\rmn{dyn}}}) 	
	\end{equation}
This encapsulates the physical assumption that the satellite cannot
merge with the central galaxy fewer than two crossing times after it
first enters the halo.
 
When the merging time counter vanishes, we add the properties 
of the satellite to those of the central galaxy, and we remove the 
satellite from the galaxy list. We introduce a
bulge formation threshold $\fbulge$ as a free parameter, 
used to match the observed abundance of galaxies by morphology. 
Mergers in which $M_{\rmn{sat}}/M_{\rmn{central}}>\fbulge$ are
considered major mergers and create an elliptical galaxy containing
the stars of both progenitors, as well as stars formed from their cold
gas component in a burst assumed to last $10^{8}$ years. A merger with
$M_{\rmn{sat}}/M_{\rmn{central}}<\fbulge$ is considered minor merger.
We add the cold gas of the satellite to that of the central galaxy.
If no bulge already exists in the central galaxy, we create a new
bulge with the stars of the merging satellite, otherwise we simply add
the stars to the existing bulge. Such a scheme is sufficient for our current
purposes, more realistic treatments include that of \citet{Som01}.  

%%%%%%%%%%%%%%%%%%%%%%%%%%%%%%%%%%%%%%%%%%%%%%%%%%%%%%%%%%%%%%%%%%%%%%%%%%%%%%%%%%%%%%%%%%%%%

\subsection{Setting parameters}
\label{sec:SA:Norm}

The models set out above have four free parameters:  the
baryon fraction $\fbar$, the star formation efficiency $\alpha$,
the supernova feedback efficiency $\epsilon$, and the bulge formation 
threshold $\fbulge$. We choose these parameters within the range which
seems physically plausible in such a way as to maximise the agreement
of the resulting galaxy populations with observation. Notice that this
procedure is only feasible in a scheme of the type we have implemented
where changes in such physical parameters do not require the
supercomputer simulations to be repeated. We attempt to match the 
following observations:
	\begin{enumerate}
	\renewcommand{\theenumi}{(\arabic{enumi})}
	\item The velocity--luminosity relation for simulated spirals
	\emph{before dust correction} should
	match the published \Iband relation of \citet{Gio97}, which
        is already corrected for internal extinction.
	\item The \bjband luminosity function should be in rough 
        agreement with those of the 2dF and SDSS galaxy surveys
        \citep{Fol99, Bla01}. Note that our poor
        resolution prevents meaningful comparison of the faint end
        slopes so we concentrate on getting reasonable fits for 
        the brighter galaxies.
   	\item Galaxies with \Vdisk $\sim 220\: \kms$ 
	should host $\sim 10^{11}\rmn{M_{\odot}}$ of stars and a few
	times $10^{9}\rmn{M_{\odot}}$ of cold gas (c.f. K99,
        SP99). The latter includes both molecular
        and atomic gas.
	\item The morphology distribution should agree with the
	observation: this is discussed further in section~\ref{sec:SA:Norm:Morpho}. 
	\end{enumerate}		

In Table~\ref{tab:SApar} we list our estimated best parameters given these
constraints, which are described in more detail in later subsections.
Note that the baryon fraction in the \tcdm model is high and 
inconsistent with standard big--bang nucleosynthesis \citep{Bu99}, although it
matches the baryon fraction observed in massive clusters \citep{Et99}. 
Interestingly, if we adopt the preferred cooling model of SP99, we can fit the
Tully--Fisher relation with baryon fractions of order $\fbar \sim 0.08$
and $\fbar \sim 0.15$ for \lcdm and \tcdm respectively, albeit with
overbright "cD" galaxies. This same trend in the required $\fbar$ was
also found by SP99. Note that taking into account the chemical
enrichment of the hot halo gas would boost the cooling rate at late 
times, and allow the observed Tully--Fisher relation to be matched for
a lower baryon fraction. (See, for example, models {\it n} and {\it c} of
\citealt{Kam99}.) We also find that lower baryon fractions are allowed
if we assume a substantial fraction ($\sim 20 \%$) of the mass of
newly formed stars is returned to the cold gas of the galaxy, as 
expected from SNe and stellar winds (see C00). We did not include this
possibility in the models described here, as it makes more sense to
use it in combination with a full chemical enrichment scheme.

\begin{table}
\caption{Parameters of the galaxy formation models. 
$\fbar=\Omega_{\rmn{bar}}/\Omega_{0}$ is the baryon
fraction, $\fbulge$ the bulge formation threshold, 
$\alpha$ and $\epsilon$ the star formation and feedback efficiencies respectively, 
and "feedback" gives the scheme that we have used.}
\begin{tabular}{llllll}
\hline
Model & $\fbar$ & $\alpha$ & $\epsilon$ & $\fbulge$ & feedback   \\
\hline
\rule{0in}{3ex}
$\Lambda$CDM & 0.12 & 0.05 & 0.05 & 0.1 & retention \\
\hline
\rule{0in}{3ex}
$\tau$CDM & 0.2 & 0.15 & 0.03 & 0.1 & ejection \\
\hline
\end{tabular}
\label{tab:SApar}
\end{table}

%%%%%%%%%%%%%%%%%%%%%%%%%%%%%%%%%%%%%%%%%%%%%%%%%%%%%%%%%%%%%%%%%%%%%%%%%%%%%%%%%%%%%%%%%%%%%

% GALSFIG

\begin{figure*}
\begin{minipage}{180mm}
\centering
\caption{The z=0 galaxy distribution in \lcdm (left panel) and \tcdm
(right panel) cosmologies. The region shown is the same 
as in Fig.~\ref{fig:DMDistFig}. We plot all galaxies 
brighter than $M_{\rmn{B}}<-19.5$ and $M_{\rmn{B}}<-21.3$ in
\lcdm and \tcdm respectively. This criterion selects $\sim 3500$ 
galaxies in both cases. The size of the symbols scales with
the \Bband luminosity and the colours follow $B-V$ index. A high resolution copy of this Figure
can be found at http://www.mpa-garching.mpg.de/NumCos/CR/High-res/index.html}
\label{fig:GalsDistFig}
\end{minipage}
\end{figure*}

%%%%%%%%%%%%%%%%%%%%%%%%%%%%%%%%%%%%%%%%%%%%%%%%%%%%%%%%%%%%%%%%%%%%%%%%%%%%%%%%%%%%%%%%%%%%%
 
In Fig.~\ref{fig:GalsDistFig} we show the present-day 
distribution of galaxies in our \lcdm model for this choice of
parameters. Galaxies are overplotted on the dark matter distribution
using symbols whose size increases with luminosity and whose colour
represents $B-V$ index. The geometry of the slice is the same
as in Fig.~\ref{fig:DMDistFig}. Note the predominance of red 
galaxies in clusters and that of blue galaxies in the field.
Notice also that the local voids are indeed nearly empty of galaxies.

%%%%%%%%%%%%%%%%%%%%%%%%%%%%%%%%%%%%%%%%%%%%%%%%%%%%%%%%%%%%%%%%%%%%%%%%%%%%%%%%%%%%%%%%%%%%%

\subsubsection{Tully--Fisher relations}
\label{sec:SA:Norm:TF}
 
In K99 the Tully--Fisher relation was plotted for central galaxies
with `spiral morphology' assuming the disk rotation velocity \Vdisk
to be equal to \Vvir the circular velocity of the host halo.
When we used the same prescription, we found a number of
low luminosity galaxies, off by more than half a magnitude both from
the observed relation and from the mean relation for the simulated
galaxies. Similar outliers can be seen in fig.~6
of K99 -- at \Vvir $\sim 300\:\kms$ some of their `Sb/Sc' galaxies are
too faint by more than one magnitude. This problem arises because
\Vvir$\sim$\Vdisk is a poor approximation for a subset of galaxies
with unusual halo assembly histories. To get a better 
estimate for \VdiskCom we proceed as follows. Throughout the evolution
of our galaxy population we accumulate $\Delta M_{\rmn{stars}}\VvirD$
for each star-forming central disk where \Vvir is the circular
velocity of its current halo. The circular velocity of the disk at any
redshift is then defined by
	\begin{equation}
	\label{eqn:Vdisk}					
	\VdiskD = {{\sum_{z_{i}}\Delta M_{\rmn{stars,z=z_{i}}}\VvirD_{\rmn{z=z_{i}}}}\over{M_{\rmn{stars,disk}}}},
	\end{equation}
where the sum is over all redshift intervals since the last
major merger and $M_{\rmn{stars,disk}}$ is the stellar mass of the disk.	
Satellite galaxies keep the disk rotation velocity they had when
they last were central galaxies.

\begin{figure*}					
	\begin{minipage}{160mm}	
	\centering	
	\epsfig{file=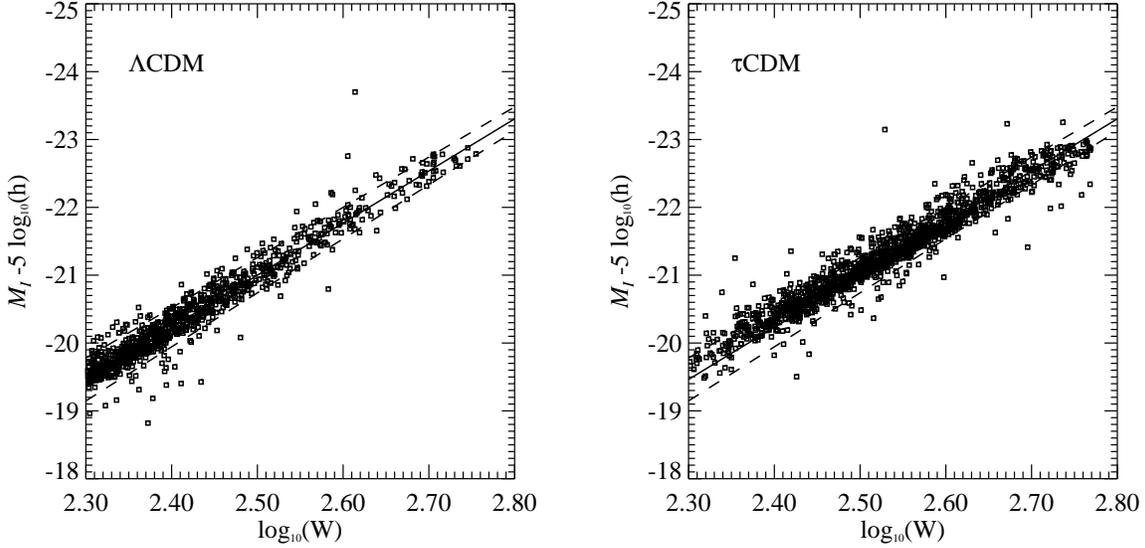,width=16cm,height=8cm}						
	\caption {Tully--Fisher relation for Sb/Sc galaxies in the simulations.  	
	The solid line is the relation which \citet{Gio97} fit to
	their observational data. The dashed lines give
        their estimate for the "intrinsic scatter" of the data.}
	\label{fig:TFFig}
	\end{minipage}
\end{figure*}

In Fig.~\ref{fig:TFFig} we plot Tully--Fisher relations for simulated
galaxies with $100<\,$\Vdisk$<300\:\kms$ and 
$1.5 < M_{B,\rmn{bulge}}-M_{B,\rmn{total}} < 2.2$ (appropriate for Sb/Sc galaxies
according to \citealt{Si86}). This is the same criterion used by K99
except that here we do not restrict ourselves to central galaxies.
We compare our results with the fit to the observational data given
by \citet{Gio97}:
	\begin{equation}
	\label{eqn:TFGiov}					
	M_{I}-5\log(h)=-21.00\pm0.02-7.68\pm0.24(\log(W)-2.5),
	\end{equation}
where $W=2\times\VdiskDot$ The dashed lines show $\pm 1\sigma$ of what
Giovanelli et al. term the ``intrinsic scatter" of their relation. This
relation has been corrected for internal extinction so we have not
included any effects of dust in our model predictions.

Even with our revised definition of \VdiskCom there are 
still a few galaxies with low \Iband luminosity which show up
below the observed relation. We could remove the five low-luminosity
galaxies at $\log(W) \sim 2.4$ 
in \lcdm by adopting a more stringent morphological selection
criterion, for example $1.6 < M_{B,\rmn{bulge}}-M_{B,\rmn{total}} < 2$. 

We have checked that the galaxies lying below (more than 0.7
magnitude) the expected relation are redder than 
"typical" spirals: their mean $B-V$ index is 0.9 and 1.0 in \lcdm
and \tcdm respectively, instead of the 0.7 found for ``normal''
spirals. These galaxies are older systems, for example,
halo satellites 
which still satisfy the morphological selection criterion in the
\BbandCom have kept the disk rotation velocity when they last were a
central galaxy, but where star formation has stopped
because of gas exhaustion, resulting in a relatively low
stellar mass and \Iband luminosity. They might thus correspond to
a subset of observed S0's.

\citet{Gio97} quote an intrinsic scatter of
$\epsilon_{int}=-0.28\:x+0.26$ for their Tully--Fisher
relation, with $x=\log{W}-2.6$. This is consistent with our
\lcdm model, but is exceeded by the \tcdm model at the bright
end. We find the dependence of the scatter on our
rather uncertain morphological selection 
criterion to be quite strong. A larger scatter is
expected in high-density cosmologies because of their late structure
formation, as pointed out by \citet{Bu01}. 

%%%%%%%%%%%%%%%%%%%%%%%%%%%%%%%%%%%%%%%%%%%%%%%%%%%%%%%%%%%%%%%%%%%%%%%%%%%%%%%%%%%%%%%%%%%%%

\subsubsection{Luminosity functions}		
\label{sec:SA:Norm:LF}

\begin{figure*}					
	\centering
	\begin{minipage}{160mm}		
	\epsfig{file=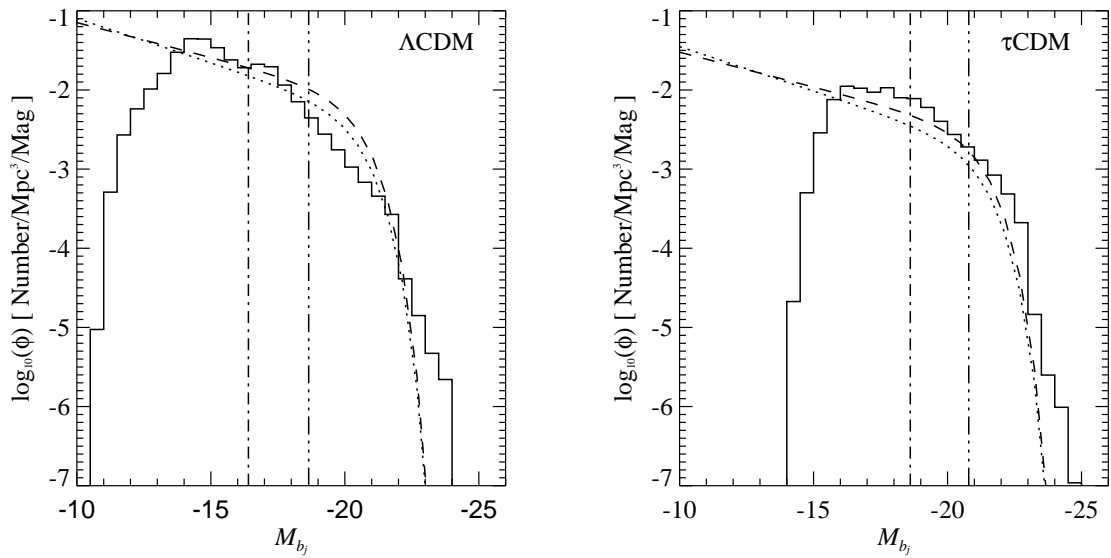,width=16cm,height=8cm}						
	\caption {\bjband luminosity functions of the simulations. 
Dashed and dotted lines show the SDSS and  2dF GRS luminosity
functions respectively.  The vertical lines show the resolution limits
	in this band. The SDSS data have been converted to the 2dF
	band as recommended by \citet{Bla01}.}
	\label{fig:LFFig}
	\end{minipage}
\end{figure*}

In Fig.~\ref{fig:LFFig} we compare the \bjband luminosity functions
(LF) of our simulations with those derived from the 2dF and 
SDSS redshift surveys. The vertical lines correspond to our 
\emph{luminosity} and \emph{morphology} resolution limits. 
As already stressed, a major difficulty in reproducing observed
luminosity functions with the kind of modelling employed here
is in establishing a sufficiently strong cut-off at high luminosities 
by avoiding the formation of overly bright central galaxies in groups 
and clusters. Despite our best efforts and our {\it ad hoc} cooling
switch at \Vvir$=350$ \kmsCom this has not been completely successfully 
achieved in the models, where we still have a bright galaxy at $M_{b_{j}} \sim -24.5$,
possibly due to the overmerging issue discussed in Section~\ref{sec:SA:Rec:Merging}. 
A second major difficulty is in suppressing the luminosity of galaxies
in low mass haloes sufficiently to reproduce the flat faint-end slope
of observed LFs. We have not stressed this issue here because our
resolution does not allow us to address it adequately.

The \lcdm LF shows a deficit of \Lstar galaxies, with respect to 
the observations, whereas the \tcdm LF produces somewhat too many
such galaxies. As pointed out by \citet{Kau93b}, this
difference is a consequence
of the differing number of $\Vvir \sim 200$ \kms haloes in the two
cosmologies, given our requirement that both should reproduce the
observed Tully--Fisher relation. The global shape of the \tcdm LF agrees
somewhat better with the observations. The shapes of both our
luminosity functions are closer to the observations than those given
in K99. This is a result of our tuning our cooling and merging
prescriptions as noted in previous sections (see also SP99 and
C00).

The results of S00 suggest that the luminosity function shape would 
be improved if we used simulations of sufficiently high resolution 
to follow the merging of galaxies directly, rather than having to use 
a phenomenological model.
The ``knee" in our simulated \bjband LFs occurs at  $M_{b_{j}} \sim -22$ in 
both our models. 93\% and 96\% of the galaxies 
brighter than these limits are central galaxies of haloes in 
\lcdm and \tcdm respectively. Of these central galaxies, 73\% and 
16\% belong to haloes with $V_{\rmn{200}} > 350$ \kmsCom showing that
our LF results in this range are being affected strongly by the
cooling cut-off and by the merging which can brighten galaxies above it.

%%%%%%%%%%%%%%%%%%%%%%%%%%%%%%%%%%%%%%%%%%%%%%%%%%%%%%%%%%%%%%%%%%%%%%%%%%%%%%%%%%%%%%%%%%%%%

\subsubsection{Masses for Milky Way look-alikes}		

We select Milky Way look-alikes as the central galaxies of
haloes with $200<$\Vdisk$<240\:\kms$ which also satisfy the ``Sb/Sc''
criterion we used when plotting the Tully--Fisher relation.
Table~\ref{tab:MWMass} presents their mean stellar and cold gas 
masses together with their \Bband and \Iband absolute magnitudes,
their colours, their disk SFR's and their abundance in the
simulations.

Our two models show similar cold gas and stellar masses, 
quite comparable with those estimated for the Milky Way, 
although their present-day star formation rates
differ by a factor of two. We find an average SFR of
1.8 \msunyr for disks of Milky Way--like galaxies in \lcdm and
3.4 \msunyr for such disks in \tcdmDot This reflects
the later formation of structure in \tcdm which requires more star
formation at late times to get enough stars
to reproduce the Tully--Fisher relation. Both values are in reasonable
agreement with the 2 to 3 \msunyr estimated for the mean star
formation rate in the Milky Way's disk over the last few Gyr 
\citep{Ro00a,Ro00b}.

The difference in the number of galaxies with 
$200<$\Vdisk$<240$ \kms in \lcdm and \tcdm is substantial but is 
fully accounted for by the difference in the halo mass functions.

%%%%%%%%%%%%%%%%%%%%%%%%%%%%%%%%%%%%%%%%%%%%%%%%%%%%%%%%%%%%%%%%%%%%%%%%%%%%%%%%%%%%%%%%%%%%%

\begin{table*}
\begin{minipage}{160mm}
\centering
\caption{Masses (in units of $\rmn{M}_{\odot}$), disk star formation rates (in
units of \msunyrKC colour and magnitudes of Milky--Way type galaxies ($200<$\Vdisk$<240$ \kmsKet}
\begin{tabular}{llllllll}
\hline
Model & Star mass  & Gas mass & disk SFR & $M_{\rmn{B}} - 5 \log h$ &
$M_{\rmn{I}} - 5 \log h$ & $B-V$ & Number \\

\hline
\rule{0in}{3ex}
\lcdm & $9.32\times 10^{10}$ & $1.05\times 10^{10}$ & 1.78 & -20.02 & -22.01 & 0.74 & 229 \\

\hline
\rule{0in}{3ex}
\tcdm & $1.58\times 10^{11}$ & $1.15\times 10^{10}$ & 3.37 & -20.04 & -21.99 &  0.71 & 941 \\

\hline
\end{tabular}
\label{tab:MWMass}
\end{minipage}
\end{table*}

%%%%%%%%%%%%%%%%%%%%%%%%%%%%%%%%%%%%%%%%%%%%%%%%%%%%%%%%%%%%%%%%%%%%%%%%%%%%%%%%%%%%%%%%%%%%%

\subsubsection{Morphologies}
\label{sec:SA:Norm:Morpho}

As in \citet{Kau93b}, we assign a morphological type based on 
the \Bband bulge--to--disk ratio: galaxies with
$L_{\rmn{B,bulge}}/L_{\rmn{B,disk}} > 1.52$ are ellipticals, S0's have 
$0.68 <L_{\rmn{B,bulge}}/L_{\rmn{B,disk}} < 1.52$, spirals have 
$L_{\rmn{B,bulge}}/L_{\rmn{B,disk}} < 0.68$ and irregulars have no bulge.

We adjust our major merger threshold parameter so that the overall
morphology fractions in our simulations match those derived by
\citet{Bau96b} from the observations of \citet{Lo96}:
13\%, 20\% and 67\% for Es, S0s, and Sps + Irrs.
We estimate these ratios in the simulations using all galaxies 
brighter than our morphology resolution limit. Taking
$\fbulge = 0.1$ we obtain 12\%, 19\% and 69\% as the relative 
fraction of the three morphological classes in \lcdmCom and 
16\%, 14\% and 70\%  in \tcdmDot For \tcdm  our S0 fraction is 
somewhat smaller than the observed 20\%.  This may be another hint 
for overmerging: central galaxies generally spend little time within
the S0 bulge--to--disk ratio limits, either growing bigger disks or
evolving into Es via mergers. Both the cooling and the merging
processes are more rapid in \tcdmCom  leading to a reduced number
of S0s. Given the different merging prescriptions used by different
authors, one must be careful when comparing the $\fbulge$ values
they derive. For example, SP99 and K99 both fit the observed morphological ratios
with  $\fbulge=0.25$, but with a different formula for the
dynamical friction timescale than is used here. As C00 emphasize, the merger
simulations of \citet{Wa96} and \citet{Ba98} suggest $0.1 \lsim \fbulge \lsim 0.3$.

\begin{figure*}					
	\begin{minipage}{160mm}	
	\centering
	\epsfig{file=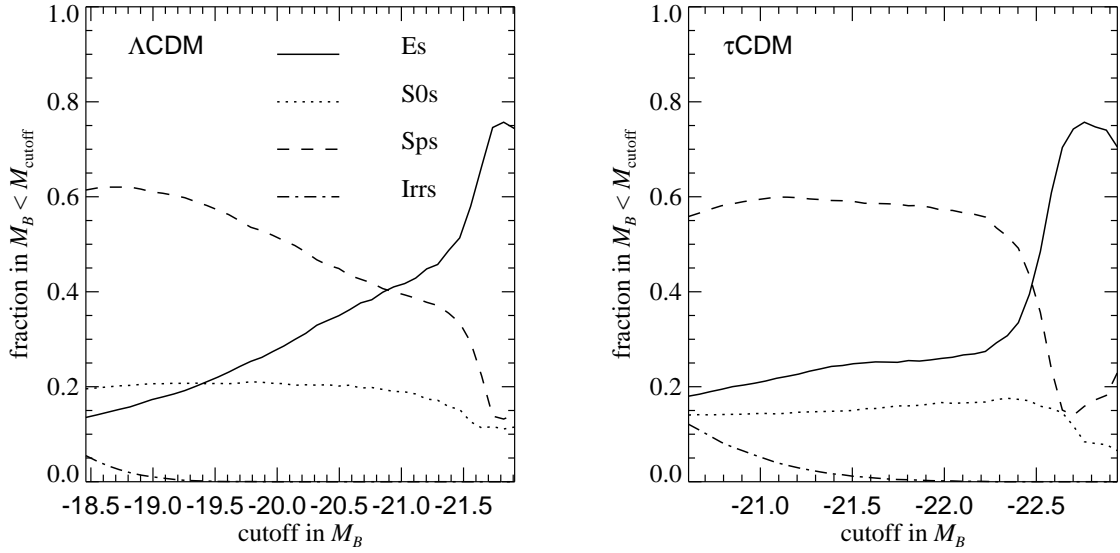,width=16cm,height=8cm}						
	\caption {Morphology distributions for galaxies brighter than a
	given absolute $M_{B}$ threshold.}
	\label{fig:AllMorphoFig}
	\end{minipage}
\end{figure*}

In Fig.~\ref{fig:AllMorphoFig}, we plot mean morphology fractions as
 a function of the maximum $B$ absolute magnitude of the
galaxies considered. The fraction of bulge--dominated galaxies
increases with luminosity. The most luminous galaxies, which are
mainly the central galaxies of clusters, are all ellipticals. The 
variation of the morphological mix with luminosity is smooth, except for
a steep rise in the elliptical fraction at $M_{\rmn{B}} \sim
-22$. This is a consequence of our {\it ad hoc} suppression of cooling
onto central galaxies in haloes with $\Vvir>350\:\kmsDot$ This suggests
that the observed predominance of  bright ellipticals in
clusters is related to the observed inability of cooling flows
to generate new disks in such systems.

\subsubsection{Colour distributions}
\label{sec:SA:Norm:Colours}

\begin{figure*}					
	\begin{minipage}{180mm}	
	\centering		
	\epsfig{file=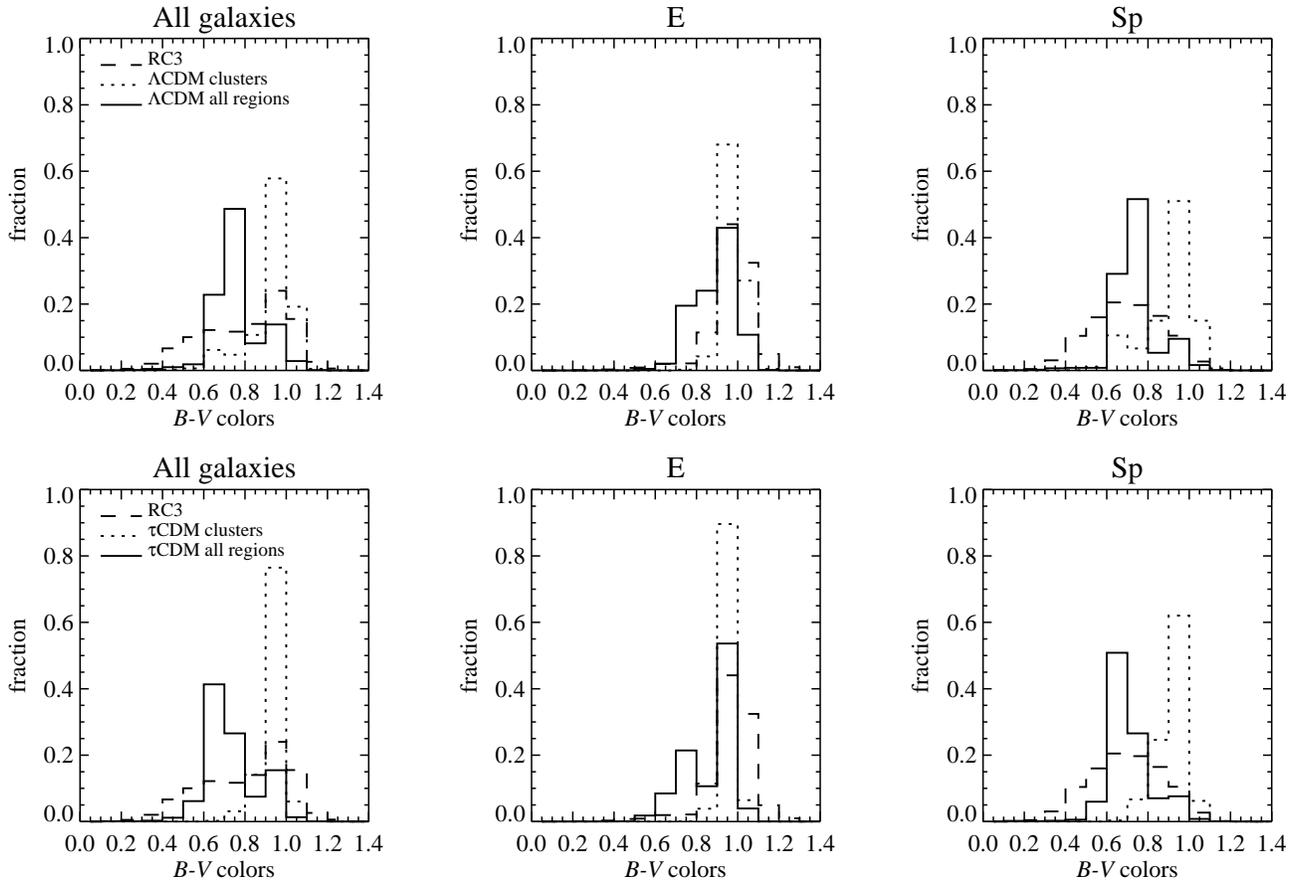,width=18cm,height=12cm}
	\caption {Colour distributions for simulated galaxies brighter
	than the morphology limit. The left panels show all galaxies,
	the central panels ellipticals, and the right panels spirals.
	The top plots are for \lcdmCom the bottom ones for \tcdmDot
	Solid histograms refer to the full simulated volume, dotted
	histograms to galaxies in clusters and the dashed histogram to
	the RC3.}
	\label{fig:ColoursFig}
	\end{minipage}
\end{figure*}

Although we do not use colours when setting the free parameters of our
models, we briefly discuss here the colour distributions of our
galaxies. The solid histogram in Fig.~\ref{fig:ColoursFig} shows 
the distribution in $B-V$ colour of all the galaxies 
brighter than our morphology limit. Two peaks are evident in these
distributions which are separated when we plot individual
distributions for ellipticals and spirals, as defined in \ref{sec:SA:Norm:Morpho}. 
There is good agreement with K99 and with the colours derived from the
RC3 catalogue of \citet{Vau91} (shown as the dashed histogram). If we restrict
ourselves to galaxies residing in massive clusters (those with $\Vvir >
500\:\kms$), we obtain the dotted plots, showing that 
cluster galaxies are redder than their field counterparts (see also
fig.~16 of S00).

%%%%%%%%%%%%%%%%%%%%%%%%%%%%%%%%%%%%%%%%%%%%%%%%%%%%%%%%%%%%%%%%%%%%%%%%%%%%%%%%%%%%%%%%%%%%%

%%%%%%%%%%%%%%%%%%%%%%%%%%%%%%%%%%%%%%%%%%%%%%%%%%%%%%%%%%%%%%%%%%%%%%%%%%%%%%%%%%%%%%%%%%%%%

\section[]{Nearby Clusters}
\label{sec:Clus}

This section begins the detailed comparison of our simulations
with the individual structures seen in the local universe.
We start with a one--to--one comparison of the largest simulated 
clusters with their nearby counterparts.
It is surprising that this comparison is
possible at all because rich clusters are systematically
under-represented in the \IRAS 1.2 Jy survey from which we
constructed our initial conditions. The next subsection shows
that a convincing identification of simulated and real clusters 
is, nevertheless, possible, and we go on in subsequent subsections
to explore this correspondance in terms of cluster 
dark matter properties, galaxy luminosity functions,
and cluster mass--to--light ratios.

%%%%%%%%%%%%%%%%%%%%%%%%%%%%%%%%%%%%%%%%%%%%%%%%%%%%%%%%%%%%%%%%%%%%%%%%%%%%%%%%%%%%%%%%%%%%%

\subsection{Observed versus simulated clusters}

Among nearby rich clusters we select Coma, Virgo, Centaurus,
Hydra and Perseus as particularly interesting for 
comparison with our simulations. All are relatively massive and lie 
within $8000 \: \kms$. We identify them with their simulated counterparts 
as follows. We take the observed coordinates and redshift from the
{\sc{nasa/ipac}} extragalactic database 
\footnote{http://nedwww.ipac.caltech.edu/forms/byname.html}, and
then look in the simulations for the most massive group less than 7
\hMpc away from the expected redshift--space position. This search
radius has been chosen arbitrarily to ensure that we recover 
all five selected clusters.

The positional agreement between real and simulated clusters, 
given in Table~\ref{tab:ClusDMProps}, is quite good except for our \lcdm
Centaurus cluster, which is 6.7 \hMpc away from the expected position. 
The very good match for Coma is particularly surprising: the
error in position is less than 2 \hMpc in both models, despite the
fact that the cluster is close to the edge of the high-resolution 
region, where the constraints are noisy due to the relatively poor
sampling of the 1.2 Jy survey.

%%%%%%%%%%%%%%%%%%%%%%%%%%%%%%%%%%%%%%%%%%%%%%%%%%%%%%%%%%%%%%%%%%%%%%%%%%%%%%%%%%%%%%%%%%%%%

\begin{table*}
\begin{minipage}{160mm}
\centering
\caption{Dark matter in nearby clusters. The first line is
\lcdmCom the second \tcdm and the last the real data. }
\begin{tabular}{@{}cccccccc@{}}
\hline

Name & SGL & SGB & cz & $M_{\rmn{Tot}}$ & $M_{\rmn{200}}$ & $R_{\rmn{200}}$ & $\sigma_{\rmn{p}}$ \\

\hhline{========}
\rule{0in}{3ex}
Coma & 90.2 & 7.10 & 7161 & 6.56 & 5.33 & 1.32 & 905,972   \\
 & 90.0 & 8.65 & 7137 & 14.8 & 11.7 & 1.72 & 1202,1261   \\
 & 89.6 & 8.32 & 6942 & -- & 7.98,4.98  & 1.64 & 821   \\

\hline
\rule{0in}{3ex}
Virgo & 107 & -11.4 & 1413 & 4.82 & 4.01 & 1.20 & 790,820 \\
 & 89.0 & 13.1 & 1221 & 2.69 & 2.55 & 1.03 & 716,729  \\
 & 102 & -3.25 & 1079 & -- & 4.58,2.04 & 1.26 & 632  \\

\hline
\rule{0in}{3ex}
Centaurus & 159 & -4.5 & 3873 & 10.4 & 4.85 & 1.28 & 982,956  \\
 & 158 & -8.0 & 3539 & 3.59 & 3.50 & 1.15 & 846,855  \\
 & 156 & -11.4 & 3298 & -- & -- & -- & 791  \\

\hline
\rule{0in}{3ex}
Hydra & 142 & -36.0 & 3417 & 4.23 & 2.59 & 1.04 & 662,709 \\
 & 132 & -39.7 & 3796 & 0.81 & 0.77 & 0.69 & 490,497  \\
 & 139 & -37.5 & 3418 & -- & 2.8,1.9 & 1.22 & 610  \\

\hline
\rule{0in}{3ex}
Perseus  & 348 & -11.8 & 4914 & 15.2 & 8.74 & 1.56 & 966,1036  \\
 & 353 & -16.5 & 5346 & 26.0 & 15.9 & 1.9 & 1293,1347  \\
 & 348 & -14.1 & 5486  & -- & 15.6,9.1 & 2.1 & 1026  \\
\hline

\end{tabular}
\label{tab:ClusDMProps}
\end{minipage}
\end{table*}

%%%%%%%%%%%%%%%%%%%%%%%%%%%%%%%%%%%%%%%%%%%%%%%%%%%%%%%%%%%%%%%%%%%%%%%%%%%%%%%%%%%%%%%%%%%%%

%%%%%%%%%%%%%%%%%%%%%%%%%%%%%%%%%%%%%%%%%%%%%%%%%%%%%%%%%%%%%%%%%%%%%%%%%%%%%%%%%%%%%%%%%%%%%

\subsection{Cluster dark matter}
\label{sec:Clus:DMProp}

In Table~\ref{tab:ClusDMProps} we list for each cluster the total 
dark matter mass (as given by our group-finder), the virial mass
(defined as the mass within \RvirKC both in units $10^{14}$ \hmsunCom
the virial radius in \hMpcCom and the line--of--sight galaxy velocity 
dispersion in units of \kmsDot We take the real data from \citet{Gir98a}. 
In the observed virial mass column, the first figure is
\MvirCom inferred from the virial theorem applied to galaxies 
within \Rvir but without including a surface term (the traditional
virial estimate). The second figure is $M_{\rmn{200,corr}}$, 
which includes a term to correct for the surface ``pressure'' and
should be more accurate (see \citealt{Gir98a} for details).
We compute line--of--sight velocity dispersions in our simulations
using all galaxies above our resolution threshold.
The first figure is obtained by considering all galaxies in 
the cluster halo, the second only those galaxies within the virial
radius. There is no significant difference. 

It is striking that the ranking of clusters by mass in the simulations
agrees with their ranking in the real universe. For individual
clusters the simulated \Mvir is usually within a factor of 2 of
the observed value and is often better. If we assume that 
$M_{\rmn{200,corr}}$ is the better observational estimator, then
for \lcdm  Coma, Hydra and Perseus agree well, but Virgo is too massive
by a factor of two, while for \tcdm  Virgo agrees well but Perseus and Coma
are too massive by a factor of two whereas Hydra is undermassive by
a factor exceeding 2. This casts some doubt on the identification of
Hydra in \tcdm -- however, it lies only $4\:\hMpc$ away from the expected
location and the next major (and more massive) cluster is $\sim 7\:\hMpc$ away.

For the line--of--sight velocity dispersion  the ranking of all five
clusters agrees with observation in both cosmologies. The individual
values also agree quite well in most cases. The larger deviations
correspond to the discrepant masses just discussed. Recall that 
Centaurus has a well--known bimodal velocity structure \citep{Lu86},
reflecting the fact that the groups Cen30 and Cen45 are currently merging
\citep{Chu99}. The quoted observed value is inferred if the two groups
are considered together. 

%%%%%%%%%%%%%%%%%%%%%%%%%%%%%%%%%%%%%%%%%%%%%%%%%%%%%%%%%%%%%%%%%%%%%%%%%%%%%%%%%%%%%%%%%%%%%

%%%%%%%%%%%%%%%%%%%%%%%%%%%%%%%%%%%%%%%%%%%%%%%%%%%%%%%%%%%%%%%%%%%%%%%%%%%%%%%%%%%%%%%%%%%%%
 
\subsection{Luminosities and M/L ratios}

\label{sec:Clus:LumProp}
	
For each cluster, we identified the most luminous galaxy
(in $B$) and calculated its $B-V$ colour. 
In Table~\ref{tab:ClusFirstRank}, we compare the results
with the data of \citet[hereafter S72]{Sa72}. For each cluster the first
line refers to the brightest cluster galaxy in our \lcdm model, 
the second to the BCG in \tcdmCom and the last to the observed
BCG. In (almost) all cases our simulated BCGs are brighter than
observed. (The BCG in the low mass \tcdm `Hydra' is an exception.) 
The discrepancy is less than about a magnitude for all objects
except the well-known peculiar galaxy NGC~1275 in Perseus, which is more
than one and a half magnitude fainter than our prediction. The observed values
quoted have been corrected for galactic extinction as follows. Most
corrections computed by S72 used a simple, latitude dependent model, 
except for Perseus where S72 estimated the correction based 
on the $B-V$ index of NGC~1275, and found $A_{V}=0.3$ and
$A_{B}=0.4$. Given the peculiar spectrum of the galaxy, these values
are quite uncertain. The \Vband magnitude of the BCG of Centaurus
found in \citet{Abe89} has not been corrected for extinction. Recall
that, owing to their low galactic latitude, 
extinction is a major issue for the \Bband
magnitude of Centaurus and Perseus. For these two clusters, we have 
therefore taken the estimation of galactic extinction given by 
\citet{Sch98} from their $100\,\mu\rmn{m}$ all-sky map of dust
emission: they find $A_{V}=0.38$ and $A_{B}=0.49$ for Centaurus
and $A_{V}=0.57$, $A_{B}=0.74$ for Perseus.

%%%%%%%%%%%%%%%%%%%%%%%%%%%%%%%%%%%%%%%%%%%%%%%%%%%%%%%%%%%%%%%%%%%%%%%%%%%%%%%%%%%%%%%%%%%%%

%%%%%%%%%%%%%%%%%%%%%%%%%%%%%%%%%%%%%%%%%%%%%%%%%%%%%%%%%%%%%%%%%%%%%%%%%%%%%%%%%%%%%%%%%%%%%

\begin{table}
\begin{minipage}{85mm}
\caption{Apparent magnitudes and colours of 
 brightest cluster galaxies. The first line is \lcdmCom the second
 one \tcdm and the last one the observations.}
\begin{tabular}{@{}llll@{}}

\hline
\rule{0in}{3ex}
Name & $m_{B}$ & $m_{V}$ & B--V \\

\hhline{====}
\rule{0in}{3ex}
Coma & 
11.57 & 10.57 & 1.00 \\
 & 11.79 & 10.81 & 0.98 \\
 & 12.50 & 11.51 & 0.99 \\

\hline
\rule{0in}{3ex}
 Virgo & 
9.64 & 8.65 & 0.99  \\
 & 8.94 & 7.94 & 1.00 \\
 & 9.42 & 8.45 & 0.97 \\

\hline
\rule{0in}{3ex}
Centaurus & 
10.47 & 9.48 & 0.99 \\
 & 10.68 & 9.68 & 1.00 \\
% & -- & 10.5 \footnote{\citet{Abe89}}  & --\\ uncorrected
 & -- & 10.12 \footnote{\citet{Abe89}}  & --\\ % corrected

\hline
\rule{0in}{3ex}
Hydra & 
11.17 & 10.18 & 0.99 \\
 & 12.42 & 11.45 & 0.97 \\
 & -- & 10.84 \footnote{\citet{Sa73}} & -- \\

\hline
\rule{0in}{3ex}
Perseus &
 10.61 & 9.62 & 0.99 \\
 & 10.79 & 9.80 & 0.99 \\
% & 12.13 & 11.54 & 0.59 \\ % corrected Sandage
 & 11.79 & 11.27 & 0.52 \\ % corrected Schlegel

\hline

\end{tabular}
\label{tab:ClusFirstRank}
\end{minipage}
\end{table}

%%%%%%%%%%%%%%%%%%%%%%%%%%%%%%%%%%%%%%%%%%%%%%%%%%%%%%%%%%%%%%%%%%%%%%%%%%%%%%%%%%%%%%%%%%%%%

The discrepancy in brightness between the simulated BCGs and
the data may be due in part to the 
luminosities of the observed BCGs being
underestimated (c.f. \citealt{Us91,Go00}), and in 
part to the overmerging problem discussed in 
\ref{sec:SA:Rec:Merging}. Note that Coma has two very bright
galaxies in a binary in its core. The luminosities of our
simulated Coma BCGs are similar to the sum of those of NGC~4889
and NGC~4874.

The colour indices of our simulated BCGs are all very red, reflecting
their old stellar populations. (Note that we use solar metallicity
population synthesis models when estimating the observed properties
of our stellar populations.) They agree well with the observed 
BCG colours in all cases except that of NGC~1275. The anomalous
A-type spectrum of this galaxy is thought to result from star
formation associated either with the strong cooling flow in Perseus
or with an apparent ongoing merger with a gas-rich galaxy.

%%%%%%%%%%%%%%%%%%%%%%%%%%%%%%%%%%%%%%%%%%%%%%%%%%%%%%%%%%%%%%%%%%%%%%%%%%%%%%%%%%%%%%%%%%%%%

\begin{figure*}					
	\begin{minipage}{160mm}	
	\centering	
	\epsfig{file=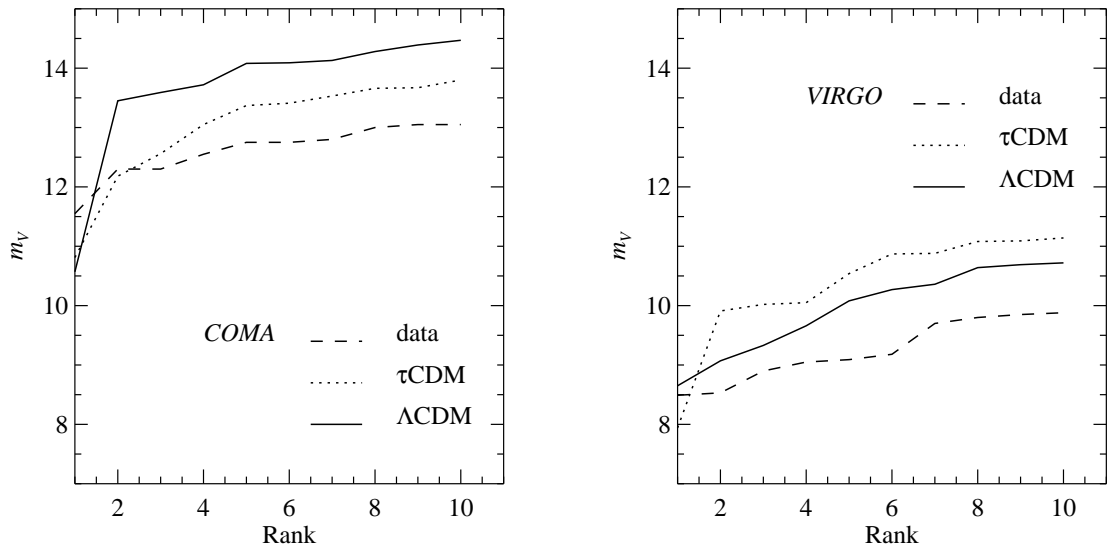,width=16cm,height=8cm}						
	\caption {Apparent magnitude in the \Vband of the brightest 
	ten galaxies in the simulated Coma and Virgo clusters compared to observations.}
	\label{fig:Clus10RankFig}
	\end{minipage}
\end{figure*}

In Fig.~\ref{fig:Clus10RankFig}, we give $m_{V}$ for the ten most luminous
galaxies in our simulated Coma and Virgo clusters. They are compared to the data
of \citet{Gu95}. The BCG  has been discussed in the previous
section, and is typically brighter in the models than in the
simulations. The remaining nine galaxies are fainter than observed
by about one magnitude. It appears that choosing parameters to fit
the Tully--Fisher relation has resulted in too few stars forming in cluster galaxies.
Possible reasons might be a systematic underestimate of \Vdisk (and thus the 
assignment of low $L$) for the central galaxies of our haloes, or an
underestimate of the effects of dust (and thus of the stellar
masses) in these same `Tully--Fisher' galaxies. The first explanation would
affect the luminosities of all galaxies and would so shift the overall
LFs of the models, the second would affect primarily cluster galaxies and
would have little effect on the global LFs. The fact that
\citet{Dia01} get a good fit (actually slightly {\it overluminous}) to the bright
end of the LF's of the CNOC1 clusters, while S00 slightly underpredict the total luminosity of Coma, suggests that the problem is not
fundamental but lies in the details of the phenomenological modelling.
The discrepancy in our own models is also evident in the cluster mass-to-light 
ratios as we now show.

%%%%%%%%%%%%%%%%%%%%%%%%%%%%%%%%%%%%%%%%%%%%%%%%%%%%%%%%%%%%%%%%%%%%%%%%%%%%%%%%%%%%%%%%%%%%%

%%%%%%%%%%%%%%%%%%%%%%%%%%%%%%%%%%%%%%%%%%%%%%%%%%%%%%%%%%%%%%%%%%%%%%%%%%%%%%%%%%%%%%%%%%%%%

\begin{table}
\begin{minipage}{140mm}
\caption{Mass--to--light ratios $\Upsilon$ of clusters.}
\begin{tabular}{@{}llllll@{}}
\hline
\rule{0in}{3ex}
 
& \multicolumn{2}{c}{\lcdm} & \multicolumn{2}{c}{\tcdm} & Data \\

\hline
\rule{0in}{3ex}
Name & $\Upsilon_{1}$ & $\Upsilon_{2}$ & $\Upsilon_{1}$ &
$\Upsilon_{2}$ &  $\Upsilon_{2}$ \\

\hhline{======}

% this for the galaxy colour Bj

\rule{0in}{3ex}
Coma  & 514 & 614  & 733 & 762  & 121-225 \\
\rule{0in}{3ex}
Virgo  & 485 & 558 & 770 & 824 & 249-459 \\
\rule{0in}{3ex}
Centaurus & 470 & 574 & 875 & 892 & -- \\
\rule{0in}{3ex}
Hydra & 484 & 657 & 629 & 651 & 162-327 \\
\rule{0in}{3ex}
Perseus & 376 & 377 & 578 & 703 & 280-458 \\

\hline
\end{tabular}
\label{tab:M2LClus}
\end{minipage}
\end{table}

%%%%%%%%%%%%%%%%%%%%%%%%%%%%%%%%%%%%%%%%%%%%%%%%%%%%%%%%%%%%%%%%%%%%%%%%%%%%%%%%%%%%%%%%%%%%%

%%%%%%%%%%%%%%%%%%%%%%%%%%%%%%%%%%%%%%%%%%%%%%%%%%%%%%%%%%%%%%%%%%%%%%%%%%%%%%%%%%%%%%%%%%%%%

% Mass-to-light ratios
 
We compare simulated mass--to--light ratios in 
the \bjband with those derived by \citet{Gir00}. 
We approximate the \bj magnitude by taking 
the ``galaxy colour relation'': $b_{j}=B-0.35\:(B-V)$ 
\citep{Bla01}. We estimate two ratios for our simulated clusters,
one using the total mass and light in our simulated haloes, the other
using values within their virial radii:
\begin{equation}
	\Upsilon_{1}={{M_{\rmn{tot}}}\over{L_{b_{j},\rmn{tot}}}},
\end{equation}
\begin{equation}
	\Upsilon_{2}={{M_{\rmn{vir}}}\over{L_{b_{j},\rmn{<Rvir}}}}.
\end{equation}
We give all ratios in solar units for $h=1$.

If $\phi_{b_{j}}$ is our luminosity function for the \bjbandCom 
we include an estimated contribution to the total luminosity 
from galaxies fainter than our luminosity resolution limit
as follows:
\begin{equation}
	L_{b_{j},\rmn{tot}}=L_{b_{j},\rmn{faint}}+L_{b_{j},\rmn{bright}}
\end{equation}
\begin{equation}
	L_{b_{j},\rmn{bright}}=\sum_{L_{b_{j}}>\rmn{resolution}}L_{b_{j},i}
\end{equation}
\begin{equation}
	L_{b_{j},\rmn{faint}}\sim{{N_{bright}}\over{\int_{L_{b_{j},\rmn{res}}}^{\infty}\:\phi_{b_{j}}(L)\:dL}}\times
	\int_{0}^{L_{b_{j},\rmn{res}}}\:L\:\phi_{b_{j}}(L)\:dL	
\end{equation}
Table~\ref{tab:M2LClus} lists the results: the two numbers cited for 
the data are the minimum and maximum values found by
\citet{Gir00} when varying their sample of galaxies, their maximum
clustercentric radius, and their method to estimate luminosities.
For Centaurus, \citet{Gir00} did not provide any data since the cluster is multipeaked.
The observed ratios of \citet{Gir00} in the \bjband scatter around a value of 250.

For our \lcdm model, we find values scattering around $\Upsilon_1\sim 500$
and $\Upsilon_2\sim 600$, and around $\Upsilon_1\sim 600$
and $\Upsilon_2\sim 700$ for the \tcdm model, although 
with a noticeably larger scatter.

The higher value within the virial radius
is a consequence of the radial variation of the morphology and colour
of the galaxies. The typical discrepancy with 
observation appears to be factor of 2 and 2-3 for the \lcdm and
\tcdm models respectively. This is somewhat larger than
expected from the results presented above since our simulated cluster
masses are not systematically in error, the brighter galaxies in
our clusters are typically too faint by a magnitude or less,
and our cluster luminosity functions $\phi_{b_{j}}$ have faint
end slopes (which are important for the luminosity correction
discussed above) which are steeper than those of most observational
determinations \citep{Fol99,Bla01}. 

Further work, including careful analysis of the
observational mass determination techniques is needed to understand 
how much of this discrepancy comes from incorrect galaxy modelling and how
much from systematics introduced by the mass and luminosity measurement 
techniques. For example, the higher M/L ratios generically
obtained in the \tcdm model compared to \lcdm are not reflected 
in the Johnson \BbandCom where the figures are similar.
At these wavelengths, our \lcdm simulation gives total M/L ratios 
of 536, 501, 512, 490 and 402 for Coma, Virgo, 
Hydra, Centaurus and Perseus respectively. Using 
modelling techniques very similar to ours, K99 found the \Bband
M/L ratios of clusters in \lcdm to be biased with respect to that 
of the Universe as a whole by an amount which is quite consistent 
with observation.  In an independent analysis \citet{Ben00b} obtained
$\Upsilon_{\rm B}\sim 400$ for massive haloes in \lcdmDot  For comparison,
\citet{Ken82} measured $\Upsilon_{\rmn{B}}=360$ for the Coma cluster, 
perhaps the best observed of local clusters. 

It is also possible to use the galaxy populations in our simulations
to study the radial distribution of galaxies within clusters. Detailed
analyses of this kind were carried out by \citet{Dia99,Dia01}
comparing their GIF simulations to CfA and CNOC clusters. At the
resolution of our (and their) simulations, it is only possible to get
useful results by stacking many clusters. We have done this with our
data and find results which are very similar to that of the earlier
work. In the mean the radial distribution of galaxies within clusters
is similar to but slightly less concentrated than that of the dark
matter. Blue and star-forming galaxies are much less concentrated
towards the cluster centres than redder galaxies. The proportion
of elliptical galaxies rises (and that of spirals drops) near the
cluster centre. We do not repeat the plots of the earlier paper here
since we have little to add. Further progress on these issues
requires simulations of significantly higher resolution, such as
those of S00.

%%%%%%%%%%%%%%%%%%%%%%%%%%%%%%%%%%%%%%%%%%%%%%%%%%%%%%%%%%%%%%%%%%%%%%%%%%%%%%%%%%%%%%%%%%%%%

\section[]{Comparison catalogues}
\label{sec:ref}

In this section, we recall briefly the features of the two local galaxy
catalogues with which we compare our simulations. We use the far-infrared
selected \PSCZ catalogue since it includes the IRAS 1.2 Jy 
survey which was used as the density constraint on our initial conditions.
In addition it has near full-sky coverage and the highest available
IR source density in the region we have modelled. A complication is 
that comparison with this survey requires us to
model the FIR luminosity of simulated galaxies. At optical wavelengths, 
we have chosen the recently completed \UZC as our reference 
catalogue. Its sky coverage is substantially smaller than that of the 
\PSCZ but it has a higher source density and
we can compare the observed \Bband luminosities directly to our
simulations.

%%%%%%%%%%%%%%%%%%%%%%%%%%%%%%%%%%%%%%%%%%%%%%%%%%%%%%%%%%%%%%%%%%%%%%%%%%%%%%%%%%%%%%%%%%%%%

\begin{table*}
\begin{minipage}{160mm}
\centering
\caption{Number of galaxies in mock and real \PSCZ and \UZC surveys.}
\begin{tabular}{@{}lllll@{}}
\hline
Model & mock \PSCZ catalogue & \PSCZ data  & mock \UZC catalogue & \UZC data \\
\hline
\rule{0in}{3ex}
\lcdm & 6806 & 6735 & 8061 & 8031 \\ 
\hline
\rule{0in}{3ex}
\tcdm & 5410 & 5412 & 7408 & 7421 \\ 
\hline
\end{tabular}
\label{tab:CatPop}
\end{minipage}
\end{table*}

%%%%%%%%%%%%%%%%%%%%%%%%%%%%%%%%%%%%%%%%%%%%%%%%%%%%%%%%%%%%%%%%%%%%%%%%%%%%%%%%%%%%%%%%%%%%%

%%%%%%%%%%%%%%%%%%%%%%%%%%%%%%%%%%%%%%%%%%%%%%%%%%%%%%%%%%%%%%%%%%%%%%%%%%%%%%%%%%%%%%%%%%%%%

\subsection{\PSCZ catalogue}
\label{sec:ref:PSCZ}

The \PSCZ catalogue has been described by \citet{Sau00}. It contains
15411 \IRAS galaxies and covers some 84\% of the sky, excluding
regions of low galactic latitude and cirrus, and regions unobserved 
by the \IRAS satellite. 1.2\% of \IRAS galaxies
within the \PSCZ region and with $b_{j} < 19.5$ have unknown redshift, 
and we discard them from the catalogue. We then consider the
\PSCZ catalogue is limited both at $b_{j} < 19.5$ and at 
$f_{60\,\mu\rmn{m}} > 0.6 $ Jy. Also, we only consider
galaxies with $cz < 8000\: \kmsDot$

To compare the results of our simulations with the \PSCZCom we need to
estimate the IR luminosity of the galaxies at  $60\,\mu\rmn{m}$. 
Our model as described so far does not provide us with thisg
information. We assume that the $60\,\mu\rmn{m}$ luminosity of a 
star-forming galaxy has two components, one coming from star-forming
regions and proportional to its star formation rate, the other 
representing the re-emission of obscured light from older stars 
and proportional to its \Iband luminosity. For very low star formation
rates we assume a negligible FIR luminosity. This threshold in
star formation rate is set empirically. Formally, we write:
\begin{equation}
L_{\rmn{FIR}}=a_{\rmn{SFR}}\:\dot{M}_{*}+\alpha_{\rmn{I}}\:L_{\rmn{I}}\quad\rmn{
\: if }\quad \dot{M}_{*} > 0.01\:{{M_{*}}\over{t_{\rmn{Hubble}}}} 
\end{equation}
\begin{equation}
L_{\rmn{FIR}}=0  \quad \rmn{ \: otherwise}	
	\label{eqn:Lfir}	
\end{equation}
This is known to be at best a rough approximation for \IRAS
galaxies \citep{Hel86}, but it gives sufficiently accurate
results for the purposes of this
paper. A more detailed analysis would require consistent 
inclusion of the IR band in our SA model 
(see \citealt{De00}). We determine the proportionality coefficients 
$a_{\rmn{SFR}}$ and  $\alpha_{\rmn{I}}$ for each cosmology by
matching the IR luminosity function of simulated galaxies brighter than the
resolution limit to the \IRAS luminosity function given in
\citet{Sau90}. Because of resolution effects
we are able to fit the observed \IRAS luminosity
function only above a limiting $60\,\mu\rmn{m}$ luminosity of 
$L_{\rmn{60}}=2.7\:10^{8} \rmn{L}_{\odot}$ and $L_{\rmn{60}}=1.50\:10^{8}
\rmn{L}_{\odot}$ in \lcdm and \tcdm respectively, which we define as 
our FIR resolution limits. Here, $L_{\rmn{60}}$ is expressed in units of
the bolometric luminosity of the sun, and is consistent with the
assumed galaxy FIR spectrum of \citet{Sau90}. To avoid confusion we will in the
following refer to our previous luminosity resolution limits as
the ``optical resolution limits" to distinguish them from the
``FIR resolution limits". Interestingly, the fitting values of $a_{\rmn{SFR}}$ that we obtain: 
$a_{\rmn{SFR}}=1.68\:10^{9}$ and $a_{\rmn{SFR}}=6.4\:10^{8}$ in 
\lcdm and \tcdm respectively are close to the  
one estimated by \citet{Row00b} for bright IR galaxies. 
As a check, we compare the number of simulated and observed galaxies 
with $f_{{60\,\mu\rmn{m}}} > 0.6 $ Jy, $cz < 8000\:\kms$,  
$b_{j} < 19.5$ and brighter than both the optical and FIR resolution limits. 
The figures are reported in Table~\ref{tab:CatPop}. 
Note that they are different for our two cosmologies because of the 
differing simulation resolution limits.

\subsection{\UZC catalogue}
\label{sec:ref:UZC}

According to \citet{Fal99}, the \UZC is $\sim 98\%$
complete for $m_{\rmn{Zwicky}} \lid 15.5$ in the ranges
$\rmn{20^{h}} \lid \alpha_{\rmn{1950}} \lid \rmn{4^{h}}$ and 
$\rmn{8^{h}} \lid \alpha_{\rmn{1950}} \lid \rmn{17^{h}}$, for
declinations $-2^\circ.5 \lid \delta_{\rmn{1950}} \lid 50^\circ$. This
covers roughly one third of the sky. 
We again consider only galaxies with $cz < 8000\:\kmsCom$ and with
absolute magnitudes above the optical resolution limit for 
our simulated galaxies.

To compare with the \UZCCom we need to cut the simulated galaxy 
catalogues in the appropriate \Bband 
magnitude. However, the assumption of $M_{Zwicky} \sim
M_{B}$ is known to be rather approximate \citep{Hu76,Bo90}, with
a $1\:\sigma$ error of 0.3 mag. We adjust our threshold in simulated
$m_{B}$ to obtain good agreement between the predicted and observed
number of galaxies in the \UZC region. This process is independently 
applied to both our models. It leads us to adopt $M_{Zwicky} -
M_{B} = -0.4$ and $+0.9$ for the \lcdm  and \tcdm  models 
respectively. These shifts are larger than can plausibly be attributed
to the difference in photometric systems, but allow us to compensate
in part for the systematic offsets in luminosity function that are
visible in Fig.~\ref{fig:LFFig}. Again, the resulting number of
galaxies in our real and mock \UZC catalogues are reported in
Table~\ref{tab:CatPop}.

%%%%%%%%%%%%%%%%%%%%%%%%%%%%%%%%%%%%%%%%%%%%%%%%%%%%%%%%%%%%%%%%%%%%%%%%%%%%%%%%%%%%%%%%%%%%%

%%%%%%%%%%%%%%%%%%%%%%%%%%%%%%%%%%%%%%%%%%%%%%%%%%%%%%%%%%%%%%%%%%%%%%%%%%%%%%%%%%%%%%%%%%%%%

\section[]{Smoothed density fields}
\label{sec:Smooth}

The constrained density field used to set up the initial
conditions for our two simulations was derived from the \IRAS
1.2 Jy catalogue, smoothed with a Gaussian of dispersion 5 \hMpcDot
As a result we expect the galaxy density fields in our simulation to
resemble each other and to resemble the real galaxy field when
smoothed on similar scales. In this section we will check how well
this works out. In the following, we will denote by ``optical galaxies" 
the sample of all simulated galaxies brighter than our \Bband resolution
limit. Similarly, we will call ``FIR galaxies" all simulated galaxies brighter
than both the optical and the FIR resolution limits. We will apply
{\it apparent} magnitude limits to these samples only in section
6.4 where we compare with the observed \PSCZ  sample.

%%%%%%%%%%%%%%%%%%%%%%%%%%%%%%%%%%%%%%%%%%%%%%%%%%%%%%%%%%%%%%%%%%%%%%%%%%%%%%%%%%%%%%%%%%%%%

\subsection{Smoothed maps}
\label{sec:Smooth:Smoothing}

\begin{figure*}					
	\begin{minipage}{160mm}	
	\centering	
	\epsfig{file=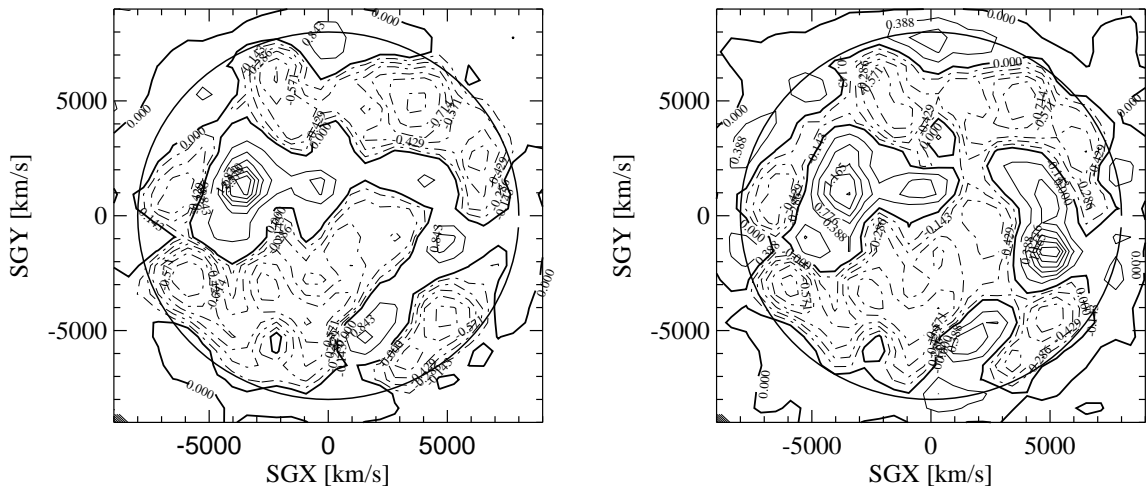,width=16cm,height=7.181cm}						
	\caption {Isodensity contours in the Supergalactic Plane of
	the distribution of simulated galaxies brighter 
	than the optical resolution limit after smoothing with a Gaussian of
        dispersion 5 \hMpcDot \lcdm
	is on the left, \tcdm on the right.}
	\label{fig:CRSmoothOnSG5MpcFig}
	\end{minipage}
\end{figure*}

\begin{figure*}					
	\begin{minipage}{160mm}	
	\centering	
	\epsfig{file=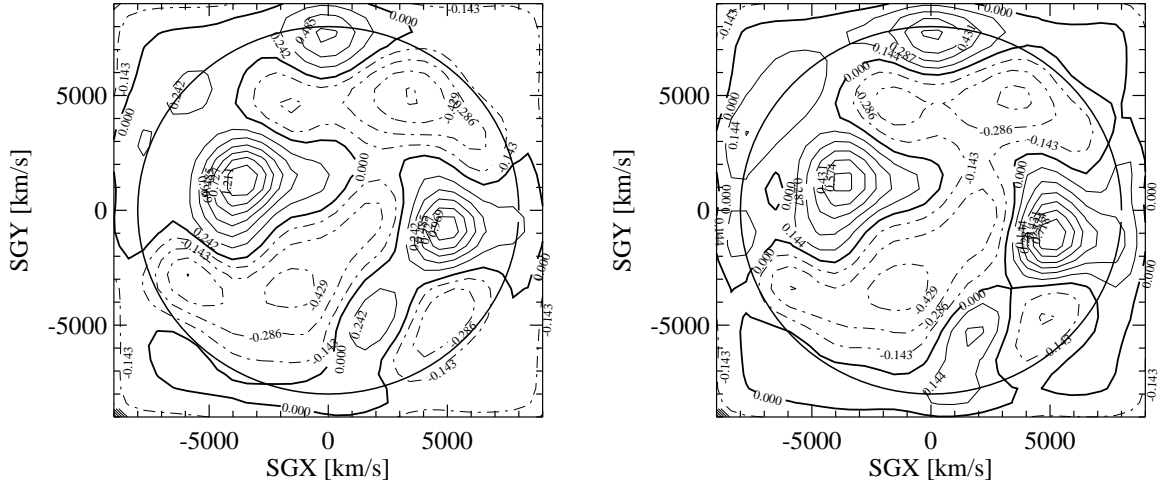,width=16cm,height=7.181cm}						
	\caption {Isodensity contours as in Fig.~\ref{fig:CRSmoothOnSG5MpcFig}
	except for a smoothing of 10 \hMpcDot}
	\label{fig:CRSmoothOnSG10MpcFig}
	\end{minipage}
\end{figure*}

As a qualitative example, Figs.~\ref{fig:CRSmoothOnSG5MpcFig}
and~\ref{fig:CRSmoothOnSG10MpcFig} show the overdensities of 
optical galaxies in the supergalactic plane, using Gaussian
kernels with 5 and 10 \hMpc smoothing length
respectively (the symbol $R$ in Equation~\ref{eqn:SmoothKernel} below). 
As expected, there is already good agreement between the
two cosmologies on the 5 \hMpc smoothing
scale, and the match is even better at 10 \hMpcDot The three major 
overdensity peaks are Coma at the top of the plot, 
the Great Attractor at the center--left,
and the Pisces-Perseus complex at the center--right. The Virgo
cluster/Local Supercluster complex is also evident just above the
centre of the plots. These maps can be compared with galaxy density
maps reconstructed directly from various \IRAS surveys
(e.g. \citealt{Br99}). The resemblance is quite close and is 
encouraging because of the many steps 
between using the observed density fields 
to constrain the initial dark matter distributions in our models 
and constructing simulated galaxy density fields from the galaxies 
which are formed during their evolution.

%%%%%%%%%%%%%%%%%%%%%%%%%%%%%%%%%%%%%%%%%%%%%%%%%%%%%%%%%%%%%%%%%%%%%%%%%%%%%%%%%%%%%%%%%%%%%

%%%%%%%%%%%%%%%%%%%%%%%%%%%%%%%%%%%%%%%%%%%%%%%%%%%%%%%%%%%%%%%%%%%%%%%%%%%%%%%%%%%%%%%%%%%%%

To obtain more quantitative results, we now compare various smoothed
density fields point-by-point. To do this we assign the dark matter
particles and the galaxies to a regular grid with a spacing of 10 \hMpc 
using a CIC scheme. We keep only those grid points within $cz<8000\: \kmsCom$
resulting in some $\sim 2100$ nodes. We further smooth this density
field using a Gaussian kernel of the form
\begin{equation}
	\label{eqn:SmoothKernel}
	W(r)={{1}\over{(2 \pi R^{2})^{3/2}}}\exp{{-{r^{2}}\over{2R^{2}}}}
\end{equation}
We take $R=10$\hMpcCom corresponding to twice the smoothing used when
generating the initial conditions. 
After this smoothing, we correct for artificial effects introduced by
the spherical boundary of our galaxy distribution. This is done by
multiplying the smoothed field by the ratio of
fields obtained by applying similar smoothing procedures to two
Poisson distributed catalogues, one uniform within a box of side
$220\:\hMpc$ centred on the simulation sphere, the
second uniform within the region where our data lie
($cz<8000\: \kmsCom$ together with the \PSCZ mask when
appropriate). We also evaluate 
the volume of the intersection between the smoothing kernel and 
our selection region using this simple Monte-Carlo scheme and
we discard nodes where this volume is less 
than half of the volume of the smoothing kernel.

Note that we have chosen the spacing of our grid so that the smoothed
density fields are not heavily oversampled, while retaining a reasonable
number of data points. For consistency, we define the mean density
of each smoothed field as the straight average of its values at
our final set of nodes. We can then define the smoothed overdensity at
each node through		
\begin{equation}
		\rmn{\delta_{\rmn{s,10}}}={{\rmn{\rho_{\rmn{s,10}}}-1}\over{\overline{\rho}_{\rmn{s,10}}}}
\end{equation}

We now discuss scatter plots which show point-by-point comparisons 
of various of these density fields. We begin by focussing on different
components within a given simulation. We then compare  each component
between our two simulations. Finally we compare our simulated FIR
galaxy density fields with the real \PSCZ data.

%%%%%%%%%%%%%%%%%%%%%%%%%%%%%%%%%%%%%%%%%%%%%%%%%%%%%%%%%%%%%%%%%%%%%%%%%%%%%%%%%%%%%%%%%%%%%

\subsection{Optical galaxies vs. FIR galaxies vs. mass}
\label{sec:Smooth:OptPSCZ_Mass}

\begin{figure*}					
	\centering
	\begin{minipage}{147mm}	
	\centering	
	\epsfig{file=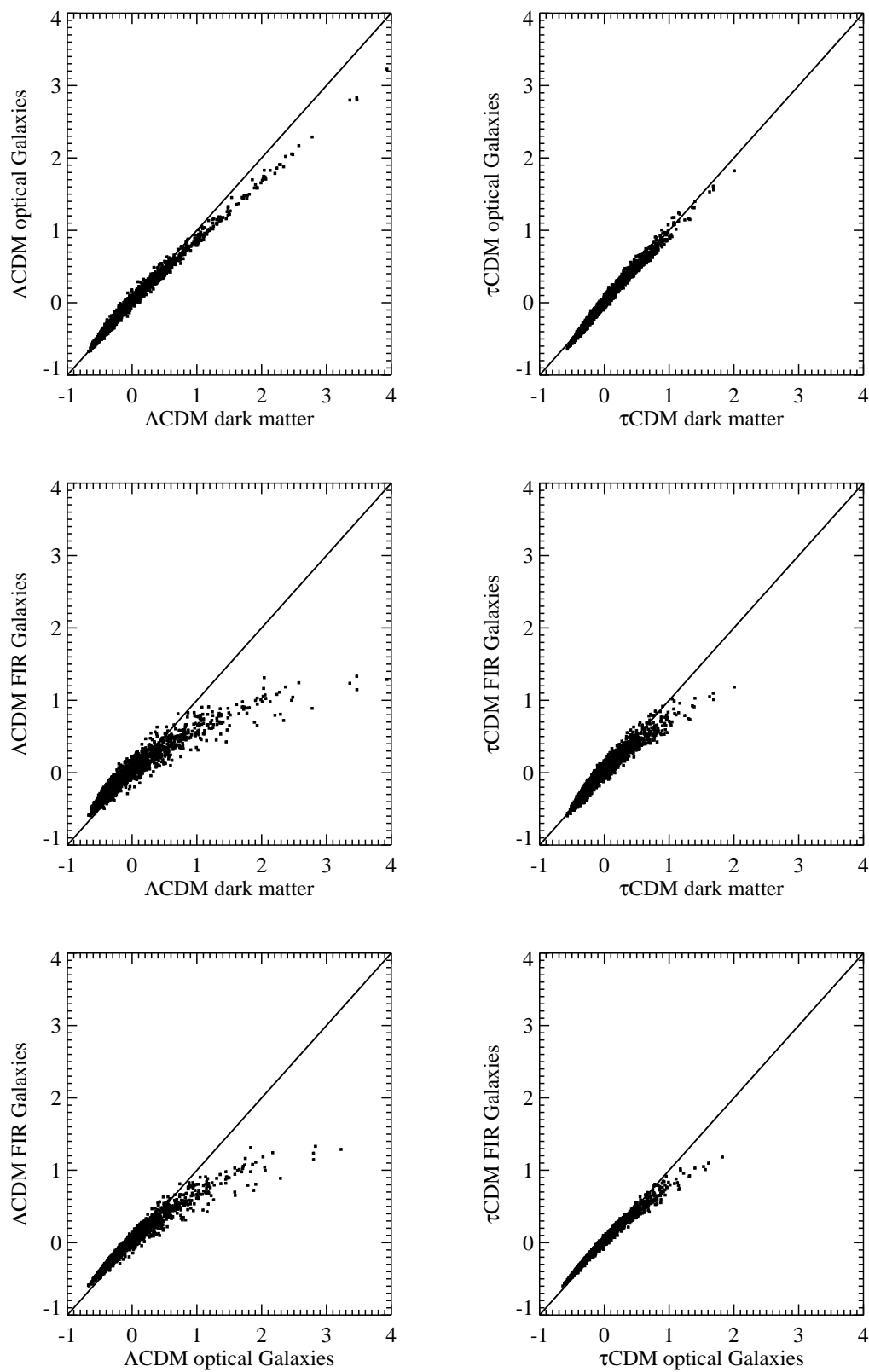,width=14.7cm,height=22cm}						
	\caption {Scatter plot comparisons of the overdensity fields
	for optical galaxies, for FIR galaxies and for mass in our two
        simulations. The top row plots optical galaxies against mass,
        the middle row FIR galaxies against mass, and the bottom row
	FIR galaxies against optical galaxies. \lcdm is on the left
	and \tcdm is on the right. The diagonal line is $y=x$ in each plot.}
	\label{fig:ScatterPSCZDM_OptDMFig}
	\end{minipage}
\end{figure*}

In Fig.~\ref{fig:ScatterPSCZDM_OptDMFig}, we compare the 
overdensity distributions of optical galaxies, of FIR galaxies and of
mass in each of our simulations.

The overdensities of optical galaxies and of dark matter are
remarkably tightly correlated in both simulations. In each case
the galaxies are almost unbiased with respect to the
underlying DM except in the highest density regions 
($\delta_{\rmn{s,10}} \sim 1 \;\rmn {to} \; 3$), where the optical 
\lcdm galaxies are significantly antibiased.

Biases are more evident and the scatter is larger when the FIR
overdensity fields are compared with those for the dark matter.
For \lcdm the trends are similar to but stronger than those for
the optical galaxies. At DM
overdensities of order $\delta_{\rmn{s,10}} \sim 3$, galaxies are 
antibiased by a factor of almost 3. This is expected given our model
for the FIR luminosity: cluster galaxies have low star formation
rates and so are assigned small or zero $L_{\rmn{FIR}}$ (consistent
with observations of most E/S0 galaxies) and so are almost all
excluded from our simulated FIR catalogues. This substantially reduces
the number of galaxies counted in the densest regions.
On the other hand, in mean density and underdense regions
star--forming spirals predominate, and these are included in both 
the optical and the FIR samples. As a result the scatter plots for the
two populations are quite similar in this density range. 

A similar but weaker pattern is observed for the \tcdm FIR galaxies
and for the same reasons. The weakening in this case reflects the fact
that the present day star formation rate in the \tcdm cosmology is
five times that in the \lcdm cosmology, and a significant number
of galaxies in dense regions have fallen in recently enough
for their ongoing star formation to take them above our 
$60\,\mu\rmn{m}$ resolution limit.
  
The larger scatter in the FIR plots may well be a real effect,
reflecting, perhaps, the stochasticity introduced by the fact that
FIR luminosity is linked to star formation rate rather than stellar mass. The
resulting underweighting of all group and cluster environments
introduces scatter as well as bias into the relation between smoothed mass and
smoothed light. We checked that the lower scatter in the optical case is not
a result of the larger number of galaxies in our simulated optical
samples. Randomly picking one third of the optical galaxies and using this
subset to construct the overdensity field does not lead to a
noticeable increase in the scatter in the optical plot.

As a consistency check, we plot in the two bottom panels of 
Fig.~\ref{fig:ScatterPSCZDM_OptDMFig} the overdensities of FIR
galaxies against those of optical galaxies. 
The effect of our FIR modelling is again very clear: in both models,
but particularly in the \lcdm model, significantly 
overdense regions in terms of optical galaxies become moderately
overdense in terms of FIR galaxies.

%%%%%%%%%%%%%%%%%%%%%%%%%%%%%%%%%%%%%%%%%%%%%%%%%%%%%%%%%%%%%%%%%%%%%%%%%%%%%%%%%%%%%%%%%%%%%

\subsection{\lcdm vs. \tcdm}
\label{sec:Smooth:LCDM_TCDM}

In this section, we compare the distributions of dark matter,
of optical and of FIR galaxies between our two cosmologies. Recall
that although the initial conditions of the two simulations are nearly
identical when smoothed
on scales of 5 \hMpc (apart from a difference in fluctuation amplitude)
they differ on smaller scales. As a result, the
build-up of galaxies is almost uncorrelated between them.

\begin{figure*}					
	\centering
	\begin{minipage}{147mm}	
	\epsfig{file=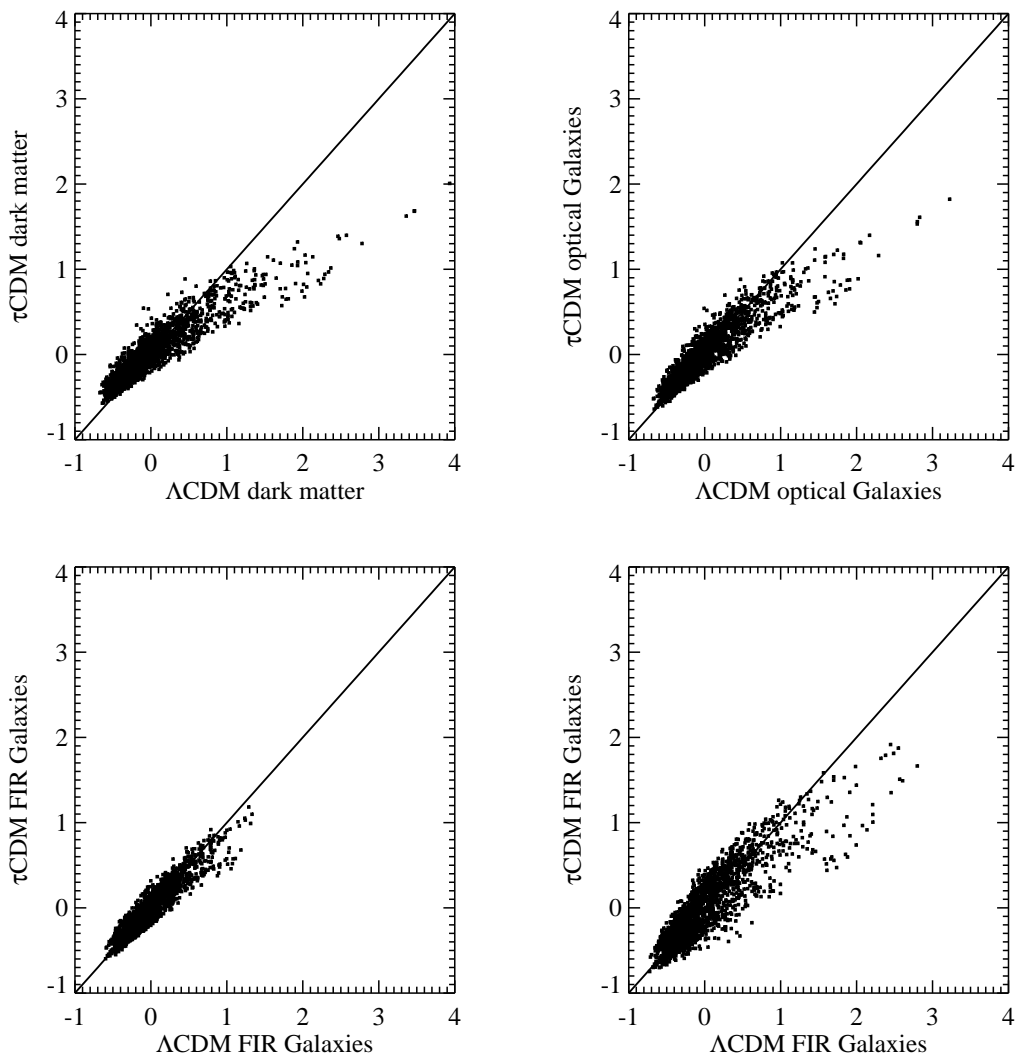,width=14.7cm,height=14.7cm}						
	\caption {Scatter plots of the overdensities 
	in \tcdm against those in \lcdmDot The upper left plot shows the dark
	matter, upper right and lower left the optical and FIR
	galaxies in real--space, lower right the FIR
	galaxies in redshift--space.}
	\label{fig:ScatterSimusTCDM_LCDMFig}
	\end{minipage}
\end{figure*}

The upper left plot of Fig.~\ref{fig:ScatterSimusTCDM_LCDMFig} compares
dark matter overdensities in the two models. The shape is as
expected from the initial set-up of the simulations. Initial amplitudes
were chosen so that the abundance of objects of rich cluster mass
would agree in the two cases. This requires the amplitude of {\it 
linear} overdensity fluctuations to be larger in \lcdm  than in \tcdmDot
The difference in amplitude carries over to the final time and shows
up as a clear ``antibias'' of the \tcdm density relative to the
\lcdm field. On the other hand, the correlation between the two
overdensity fields is strong, showing that the differences in the
initial conditions (and so in the evolved mass fields) on small
scales have little influence on larger scale fluctuations for the 
10 \hMpc  smoothing used to make these plots.

The upper right plot shows that the galaxy formation recipes we have 
	implemented, which are tuned to reproduce the observed Tully--Fisher relation
and, to a lesser extent the observed luminosity function, are only
partially successful in compensating for this difference in dark
matter clustering. The distribution of optical galaxies in \tcdm is
still significantly antibiased relative to that in \lcdmCom although the
effect is weaker than for the dark matter.

Somewhat surprisingly, this difference in clustering amplitude is
almost absent in the simulated FIR galaxy distributions. The lower
left plot in Fig.~\ref{fig:ScatterSimusTCDM_LCDMFig} shows that the 
overdensity fields of the two simulations are very similar for this
population. The higher star-formation rates of non-cluster
galaxies in the \tcdm model clearly boost the relative amplitude of
the density fluctuations in FIR population sufficiently to compensate
for the weaker clustering of galaxies by optical luminosity (or
stellar mass).

In the lower right plot, we compare the \emph{redshift--space}
distributions for this same FIR galaxy population.  As expected, the 
scatter is greater in redshift space and the overall amplitude of
the density fluctuations is increased. In addition, an extended plume 
appears at high densities. These points correspond to the environments of
massive clusters, in particular the Great Attractor and Pisces-Perseus
regions. It is interesting that peculiar velocities appear to enhance
the density contrast of these structures more strongly in \lcdm than
in \tcdmDot

From this series of four plots we conclude that the point-by-point 
agreement between our two simulations is quite good for all three 
components, at least for the rather large smoothing employed here. 
As expected, the mass is more weakly clustered in \tcdmCom and this effect
carries over to the optical galaxy population. The smoothed FIR galaxy
populations in the two simulations are very similar.

%%%%%%%%%%%%%%%%%%%%%%%%%%%%%%%%%%%%%%%%%%%%%%%%%%%%%%%%%%%%%%%%%%%%%%%%%%%%%%%%%%%%%%%%%%%%%

\subsection{Simulated \PSCZ galaxies vs. observations}
\label{sec:Smooth:PSCZ_Obs}

We  now compare the distribution of the simulated FIR galaxies
directly to the observed distribution in the \PSCZ  catalogue. This
requires a slightly different FIR sample than that used in the last
section. In addition to the redshift and simulation resolution limits
already enforced, we need to apply the sky mask of the observed
catalogue and its flux limits at 0.6 Jy and $b_j < 19.5$. The 
effects of these additional selection criteria are consistently taken
into account when constructing overdensity fields both for the
observations and for the simulated data.

\begin{figure*}					
	\centering
	\begin{minipage}{147mm}	
	\epsfig{file=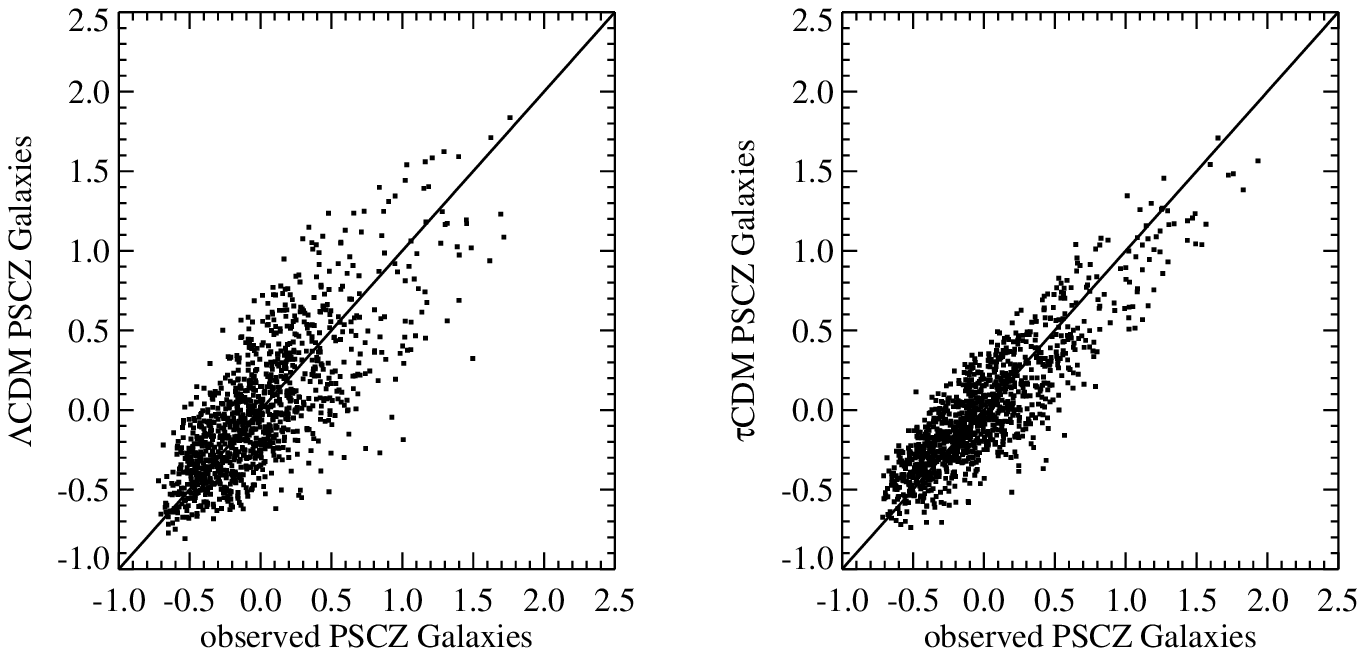,width=14.7cm,height=7.35cm}						
	\caption {Scatter plots of the redshift-space overdensities of 
        simulated \PSCZ galaxies against those of the real \PSCZ data.}
	\label{fig:ScatterPSCZDATA_PSCZFig}
	\end{minipage}
\end{figure*}

The left and right plots of Fig.~\ref{fig:ScatterPSCZDATA_PSCZFig} 
compare simulated and observed overdensity fields for the
\lcdm and \tcdm cosmologies respectively. These fields are, of
necessity, constructed in redshift--space. The agreement is reasonably
good, with greater scatter in the \lcdm case. In neither cosmology is there 
any obvious bias between simulation and observation. This is a
nice confirmation that our SA and FIR luminosity modelling schemes
produce an FIR galaxy population with a large-scale distribution
which quite closely resembles that used to define our initial
conditions.

%%%%%%%%%%%%%%%%%%%%%%%%%%%%%%%%%%%%%%%%%%%%%%%%%%%%%%%%%%%%%%%%%%%%%%%%%%%%%%%%%%%%%%%%%%%%%

%%%%%%%%%%%%%%%%%%%%%%%%%%%%%%%%%%%%%%%%%%%%%%%%%%%%%%%%%%%%%%%%%%%%%%%%%%%%%%%%%%%%%%%%%%%%%

\section[]{Correlation functions}
\label{sec:Corr} 

In this section we use correlation techniques to extend both our 
clustering measurements and our comparison with observations to
smaller scales. We begin by using auto- and cross-correlations of the
various simulated populations to quantify small-scale bias. We then
cross-correlate observed and mock catalogues to further explore how
well we have reproduced our own cosmic neighborhood.

%%%%%%%%%%%%%%%%%%%%%%%%%%%%%%%%%%%%%%%%%%%%%%%%%%%%%%%%%%%%%%%%%%%%%%%%%%%%%%%%%%%%%%%%%%%%%

\subsection{Autocorrelations}
\label{sec:Corr:Auto} 

\begin{figure*}					
	\centering
	\begin{minipage}{160mm}	
	\epsfig{file=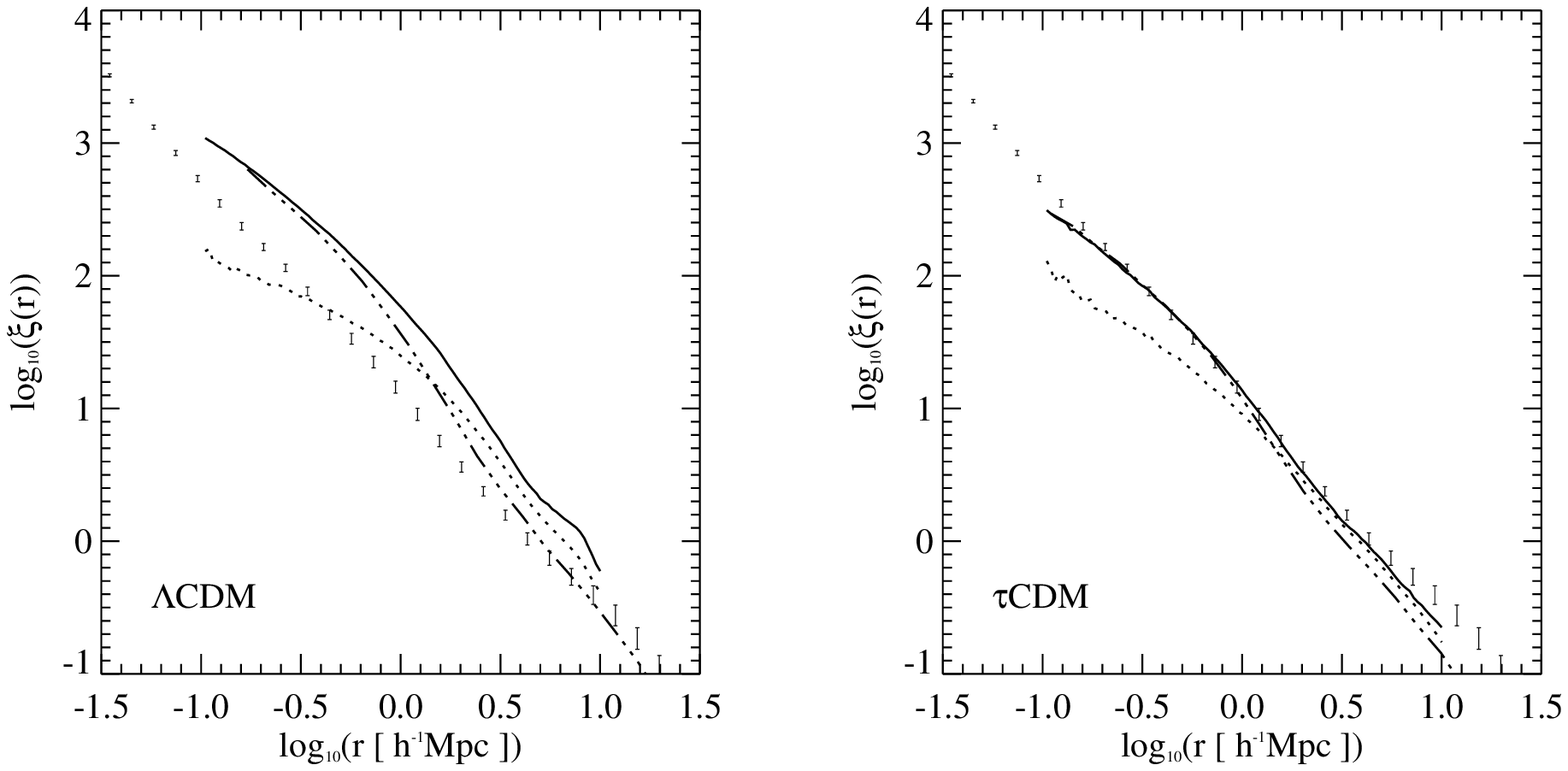,width=16cm,height=8cm}						
	\caption{Autocorrelation functions for dark matter and for
	galaxies brighter than our \Bband resolution limit.
	Solid, dotted, and dash--dotted lines are for the CR dark
	matter, the CR optical galaxies, and the \GIF dark matter 
        respectively. The error bars are the results obtained from
	angular correlations in the \APM catalogue. \lcdm is on the
	left and \tcdm on the right.}
	\label{fig:AutoCorrFig}
	\end{minipage}
\end{figure*}

In Fig.~\ref{fig:AutoCorrFig} we plot real--space autocorrelation 
functions for the dark matter and for galaxies brighter than
our \Bband resolution limit (``optical galaxies") for each of
our simulations. For reference, we also plot the dark matter
autocorrelations for the \GIF simulations analysed by K99 and the 
autocorrelation function of real optical galaxies as inferred from
inversion of the angular correlation data for the \APM survey 
\citep{Bau96c}. We compute our correlation functions up to a 
scale of 10 \hMpcDot  Note that the analysis in \citet{Je98}
shows that the \GIF results are close to the ensemble-averaged
expectations for the two cosmologies.

A comparison of the dark matter correlations between our
simulations and the \GIF simulations shows similar behaviour in 
the two cosmologies. There is reasonably good agreement on small
scales, but our simulations have more power than the \GIF models on
scales of a few Mpc. This is clearly a reflection of the particular
way in which we have constrained the large-scale density field in our
initial conditions using the observed distribution of 1.2~Jy galaxies.

On small scales the optical galaxy distribution is quite strongly 
antibiased relative to the mass in both our cosmologies, while on
scales of a few Mpc the difference is much smaller. Indeed, in \tcdm
the correlations of dark matter and optical galaxies are almost
equal beyond 1.5 \hMpcDot For neither case are the autocorrelations of
the optical galaxies close either to a power law or to the \APM data. It is
unclear whether this is a problem, since correlations calculated for 
the relatively local region we are modelling should not necessarily
reproduce those found for much larger ``representative'' regions. We
will see below that our \lcdm model does seem to reproduce the
correlation statistics of local galaxies quite well, while our
\tcdm model does not. On scales of 5 to 10 \hMpc the correlation
amplitude for optical galaxies is quite close to the \APM values in
both our simulations.

%%%%%%%%%%%%%%%%%%%%%%%%%%%%%%%%%%%%%%%%%%%%%%%%%%%%%%%%%%%%%%%%%%%%%%%%%%%%%%%%%%%%%%%%%%%%%

%%%%%%%%%%%%%%%%%%%%%%%%%%%%%%%%%%%%%%%%%%%%%%%%%%%%%%%%%%%%%%%%%%%%%%%%%%%%%%%%%%%%%%%%%%%%%

\subsection{Cross-correlations}
\label{sec:Corr:Cross} 

To make a more quantitative comparison between our simulations and the
observed \PSCZ and \UZC catalogues we have made mock \PSCZ and \UZC 
catalogues which reproduce in detail the sky coverage
and the apparent luminosity limits of the observational data. We also
limit both the mock catalogues and the real catalogues to galaxies with
absolute luminosities brighter than the relevant resolution limits of our
simulated catalogues (see above). We can then compare real and
simulated distributions in detail using auto- and cross-correlations. 
We have applied the same sky masks and depth selection functions to
our simulated mass distributions to produce mass catalogues which can
be compared directly with the observed galaxy distributions using the
same techniques.

To compute auto- and cross-correlations we use the estimator suggested 
by \citet{Ha93}:
	\begin{equation}
	\label{eqn:CrossCorr}	
	\xi_{\rmn{12}}(r)={{\langle D_{1} D_{2} \rangle \langle R_{1}
	R_{2} \rangle } \over {\langle D_{1} R_{2} \rangle \langle D_{2} R_{1} \rangle }}-1,
	\end{equation}
	where $\langle DD \rangle$, $\langle RR \rangle$ and $\langle	
	DR \rangle$ refer to the number of data-data, random-random,
	and data-random pairs, and the subscripts refer to the two
	catalogues. We compute these correlations from 0.7 \hMpc up
	to 15 \hMpcDot
        The random catalogues used here are generated
        using the same angular mask as the corresponding galaxy (or
	mass) catalogue and a selection function in depth derived from
 	the relevant luminosity function (that of \citealt{Sau90} 
	for the \PSCZCom mock \PSCZ and associated mass 
        catalogues; that of \citealt{Mar94} for the \UZC and its
	associated mass catalogue; those of~\ref{sec:SA:Norm:LF} for
        the mock \UZC catalogues). We ensure that each random
 	catalogue contains at least ten times as many points as its
	corresponding ``data'' catalogue. This ensures that
	uncertainties in pair counts are dominated by the number of
	available galaxies (or mass particles).

%%%%%%%%%%%%%%%%%%%%%%%%%%%%%%%%%%%%%%%%%%%%%%%%%%%%%%%%%%%%%%%%%%%%%%%%%%%%%%%%%%%%%%%%%%%%%

\subsubsection{Cross--correlations with FIR galaxies}

\begin{figure*}					
	\centering
	\begin{minipage}{160mm}	
	\epsfig{file=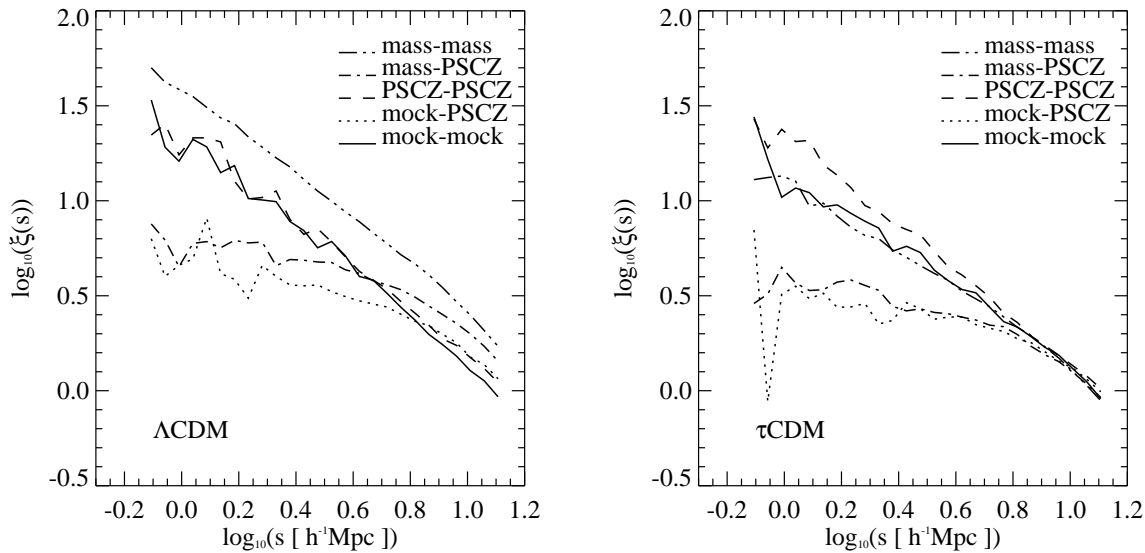,width=16cm,height=8cm}						
	\caption {Auto- and cross-correlation functions for the
	observed and simulated \PSCZ catalogues and for the mass. Note
	that the mass catalogues used here have the same sky and depth
	distribution as the observed catalogue.}
	\label{fig:CorrZ_IRASFig}
	\end{minipage}
\end{figure*}

In Fig.~\ref{fig:CorrZ_IRASFig}, we give \emph{redshift--space} auto--
and cross--correlation functions for our observed \PSCZCom mock \PSCZ and 
\PSCZ mass catalogues.

A surprising result in this figure is that the autocorrelation
of the mock \PSCZ catalogue agrees almost perfectly with that of
the real data for \lcdm and is quite close to it for separations
above 1.5 \hMpc in the \tcdm case also. In both cases the model
autocorrelation is close to a  power-law of index $-1.28$, the value 
found by \citet{Fi94} for the 1.2~Jy sample as a whole. Over the range
we plot, the bias of the simulated \PSCZ galaxies relative to the mass
is almost constant; the \lcdm galaxies are substantially antibiased
while the \tcdm galaxies are almost unbiased.

As expected, cross-correlations of the observed \PSCZ galaxies with
either our mock \PSCZ galaxies or the simulated mass distribution
show a different behaviour. On scales larger than about 5 \hMpc the
cross-correlation between real and mock galaxies is as strong as (and
effectively equal to) the autocorrelation of either population. This
is a striking confirmation of the effectiveness of our constraint
and galaxy formation procedures. On scales below 5 \hMpc this
cross-correlation flattens out to a constant value, reflecting the
fact that the small-scale structure in our simulations is unrelated to
that in the real Universe.

The cross-correlation between the observed \PSCZ galaxies and the
simulated mass distribution has a similar shape, and indeed, for \tcdm
(where the mock \PSCZ galaxies are almost unbiased) it is almost
identical to the cross-correlation with the mock galaxy catalogue.
For \lcdm the cross-correlation with mass is {\it stronger} than that
with the mock galaxies. On large scale this cross-correlation is
approximately the geometric mean of the autocorrelations of the two
populations. This would be expected for a pure linear bias model but may be 
surprising in view of the scatter and t<he nonlinearity evident in 
Figs.~\ref{fig:ScatterPSCZDM_OptDMFig} and
\ref{fig:ScatterPSCZDATA_PSCZFig}. 

%%%%%%%%%%%%%%%%%%%%%%%%%%%%%%%%%%%%%%%%%%%%%%%%%%%%%%%%%%%%%%%%%%%%%%%%%%%%%%%%%%%%%%%%%%%%%

\subsubsection{Cross--correlations with optical galaxies}
 
\begin{figure*}					
	\centering
	\begin{minipage}{160mm}	
	\epsfig{file=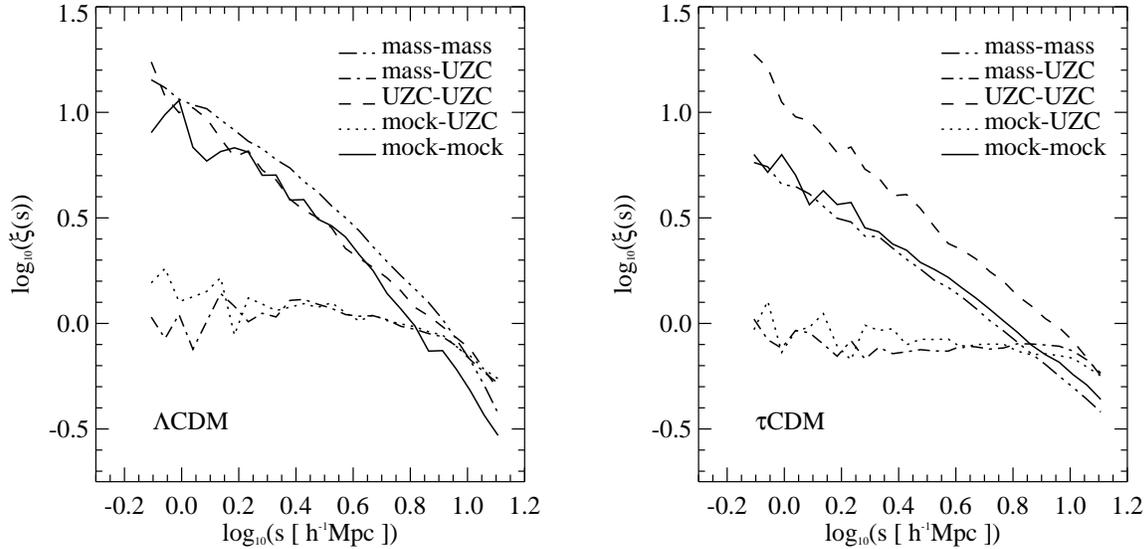,width=16cm,height=8cm}						
	\caption {Auto- and cross-correlation functions for observed
	and simulated \UZC galaxies and for the simulated mass
	distribution. \lcdm is on the left, \tcdm on the right.}
	\label{fig:CorrZ_UZCFig}
	\end{minipage}
\end{figure*}

Fig.~\ref{fig:CorrZ_UZCFig} shows auto- and cross-correlation 
functions in the same format as Fig.~\ref{fig:CorrZ_IRASFig} but
for observed and mock galaxy catalogues and for mass catalogues 
with the sky mask and depth distribution of the \UZC catalogue.
 
It is interesting to compare the autocorrelation functions of this
plot with the real-space autocorrelations in Fig.~\ref{fig:AutoCorrFig}.
As was also the case for the mock FIR galaxies, the redshift-space
autocorrelation
functions of the mock optical galaxies are parallel to those of the
dark matter; for \lcdm the galaxies are antibiased, although
less strongly so than the FIR galaxies, while for \tcdm the
galaxies show a slight positive bias. In both cases the dark matter
and galaxy curves are much more nearly parallel than are the 
corresponding real-space functions in Fig.~\ref{fig:AutoCorrFig}.
Note also the substantial differences between the mass autocorrelations
plotted in Fig.~\ref{fig:CorrZ_UZCFig} and those shown for the
same simulations in Fig.~\ref{fig:CorrZ_IRASFig}. These reflect the
different sky coverage and depth distribution of our \PSCZ and \UZC
catalogues and emphasise that neither should be considered a fair
sample of the Universe as a whole.

When we compare our mock catalogue autocorrelations with those of the
real \UZC galaxies we find excellent agreement below 5 \hMpc for
\lcdm with a slight underprediction of the observed amplitude on
larger scales. For \tcdm the predicted autocorrelations
are low on all scales, with the difference in $\xi(s)$ ranging from
30\% on large scales to a factor of 2 around 1 \hMpcDot Comparing with
Fig.~\ref{fig:AutoCorrFig} and  Fig.~\ref{fig:CorrZ_IRASFig}, we 
see that while the autocorrelations of observed optical galaxies in
our \UZC region are significantly {\it weaker} than those of observed
FIR galaxies in our \PSCZ region, they nevertheless substantially
exceed those measured for galaxies in the \APM survey (even allowing
for the difference between real and redshift space). This again
emphasises that the regions of space we are analysing are quite
small and should not be thought representative.

The cross--correlations of the observed \UZC galaxies with the
mass and mock \UZC galaxy distributions in our simulations are almost 
flat below $\sim$ 5 \hMpcCom especially in the \tcdm model. On
larger scales they do approach the autocorrelation amplitude of the
observed galaxies but the convergence is less compelling than was the
case when we compared our mock FIR galaxies with the \PSCZ data.
This difference may in part reflect the smaller volume covered by our
\UZC catalogues. It may also be due to the fact that we used
an FIR galaxy catalogue when setting up our constrained initial
conditions.

To summarise these sections on correlation functions, our initial
condition generation and galaxy evolution procedures have allowed
us to build physically consistent \lcdm and \tcdm models which
reproduce well both the individual structures and the statistical
clustering properties of star-forming galaxies in our local
neighborhood. For \lcdm the statistical properties of optically 
selected galaxies from the \UZC catalogue are also well reproduced, but
this is not the case in our \tcdm simulation where the optical
galaxies are significantly more weakly clustered.

%%%%%%%%%%%%%%%%%%%%%%%%%%%%%%%%%%%%%%%%%%%%%%%%%%%%%%%%%%%%%%%%%%%%%%%%%%%%%%%%%%%%%%%%%%%%%

%%%%%%%%%%%%%%%%%%%%%%%%%%%%%%%%%%%%%%%%%%%%%%%%%%%%%%%%%%%%%%%%%%%%%%%%%%%%%%%%%%%%%%%%%%%%%

\section[]{Mock Catalogues of Peculiar Velocities}

As an example of how our simulations can be used to test and calibrate
the  methods of cosmic-flow analysis,
we have generated Monte-Carlo mock samples mimicking real catalogues
of peculiar velocities. In particular, we show here mock catalogues
corresponding to the \MarkIII Catalogue of Peculiar Velocities
 \citep{Wi95,Wi96,Wi97}. These were made with 
an improved version of the procedure described in K96, 
which provided the basis for earlier testing of methods. 
The major advance allowed by our new simulations comes in part from
their ability to follow the nonlinear evolution of the mass distribution
on sub-Mpc scales, but more importantly from the fact that
they simulate directly the formation of the galaxies themselves.
In K96, ``galaxies'' were inserted by applying a statistical biasing
scheme to the smoothed mass density field, and their
magnitudes were drawn at random from an assumed luminosity
function. Here, galaxies form in a physically consistent way through
the condensation of gas at the centres of dark haloes and their 
morphologies reflect their actual merging histories. This more
realistic treatment should better capture possible correlations between 
galaxy properties and the underlying density and velocity fields.
Our simulations, based on constrained realizations, then allow a full  
reproduction of the correlation between systematic errors and the actual
signal of density and velocity fields.

\begin{figure}					
	\centering
	\epsfig{file=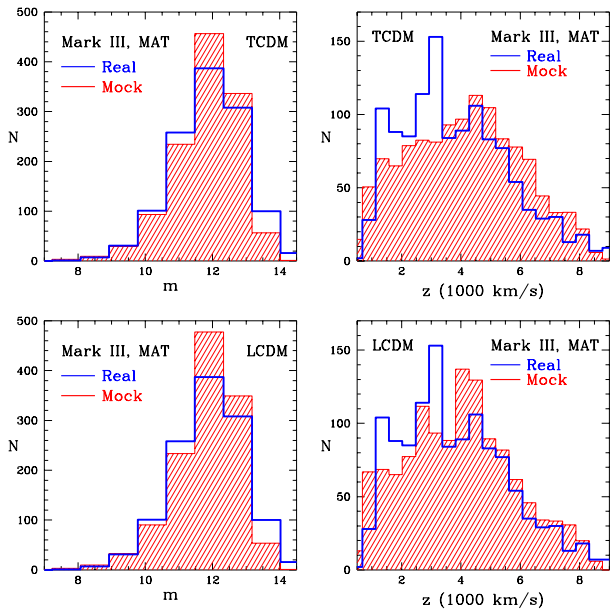,width=8cm,height=8cm}						
	\caption {Distribution of the mock apparent magnitudes and
	redshifts compared to the real data of the \MarkIII catalogue.}
	\label{fig:HistMarkIIIFig}
\end{figure}

For the purpose of this section we distinguished between ``spiral" and 
``elliptical" galaxies by
defining ellipticals as galaxies with total stellar mass 
$> 4.5 \times 10^{10} \rmn{M}_\odot $, and bulge-to-total \Vband luminosity ratio 
$ > 0.4$.
Small adjustments were made to the semi-analytic absolute magnitudes 
in order to fine-tune their match with a Schechter luminosity function.
We then assigned to each galaxy an ``observed" linewidth based on an assumed
Tully--Fisher relation and scatter as in K96 -- in practice these
linewidths were always close to those assigned by the algorithms
described earlier which are based on the actual circular velocities of
the haloes. To simulate the selection based 
on magnitude in each dataset of the \MarkIII 
catalogue, we assumed Galactic extinction as a function 
of Galactic latitude as in K96, and we slightly smeared the magnitude limits 
to take into account scatter in the relation between the magnitudes used for 
selection and those appearing in the Tully--Fisher relation.
Fig.~\ref{fig:HistMarkIIIFig} demonstrates our success in matching the final
distribution of redshift and apparent magnitude 
in our mock catalogues with that of the real data from the largest 
single dataset within the \MarkIII catalogue.

Each mock dataset was then diluted at random (trying to mimic 
selection by other independent properties such as inclination or
lack of a strong bar) to
match the number of galaxies in the real catalogues. This random
sampling, along with the random distance errors (introduced by the TF
scatter) was repeated 10 times to generate 10 mock catalogues.
The mock data were then grouped, and statistically corrected for Malmquist 
bias in exactly the same way as the real data (details are in the \MarkIII 
papers and are summarised in the \POTENT analysis, \citealt{De99}).

\begin{figure}					
	\centering
	\epsfig{file=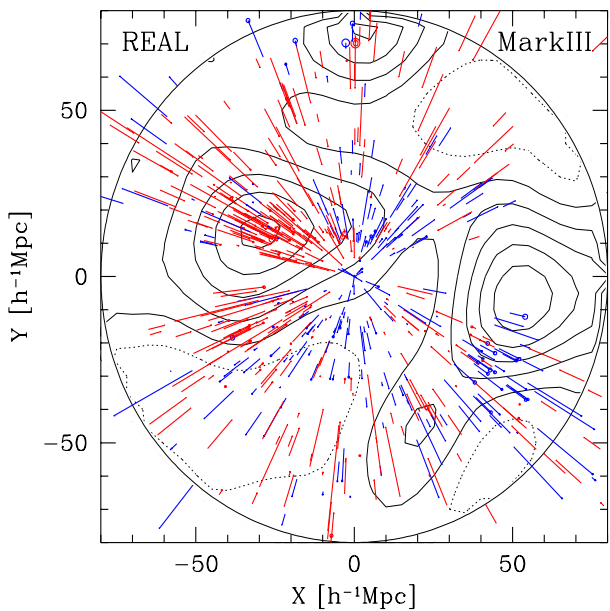,width=8cm,height=8cm}						
	\caption {The \IRAS 1.2~Jy  map of radial peculiar velocities in a slice about the 
Supergalactic plane. The contours show the density field reconstructed from
	the \IRAS data and smoothed over 12 \hMpcDot}
	\label{fig:PecVelIRASFig}
\end{figure}

\begin{figure}					
	\centering
	\epsfig{file=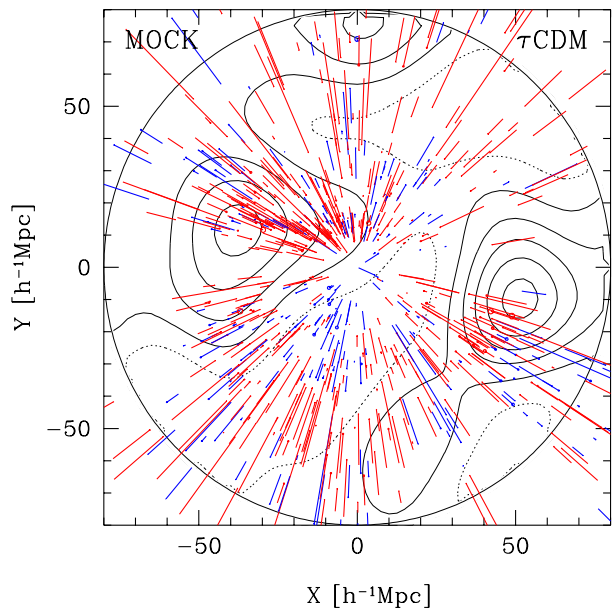,width=8cm,height=8cm}						
	\caption {Same plot as Fig.~\ref{fig:PecVelIRASFig}, but for
	the mock \tcdm catalogue. The density field shown has been
	reconstructed from the mock catalogue.}
	\label{fig:PecVelTCDMFig}
\end{figure}

\begin{figure}					
	\centering
	\epsfig{file=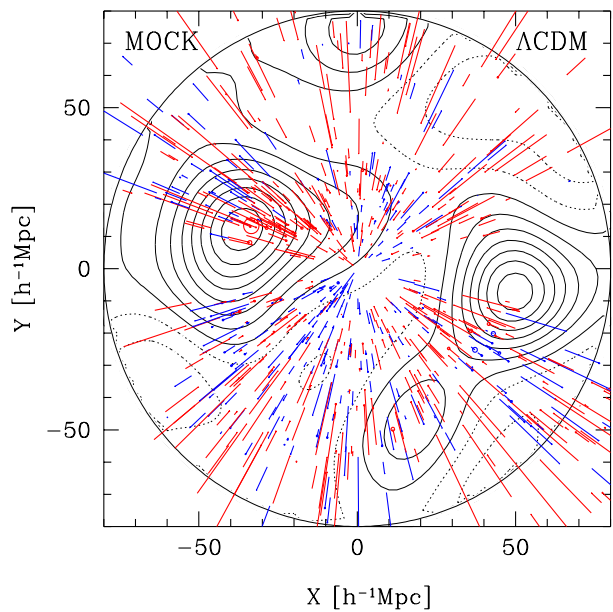,width=8cm,height=8cm}						
	\caption {Same plot as Fig.~\ref{fig:PecVelTCDMFig}, but for
	the mock \lcdm catalogue.}
	\label{fig:PecVelLCDMFig}
\end{figure}

Mock and real maps of radial peculiar velocities in a slice about the 
Supergalactic plane arEe shown in Figs.~\ref{fig:PecVelIRASFig}
to~\ref{fig:PecVelLCDMFig}, illustrating the degree of 
agreement between simulated and observed peculiar velocity fields and 
between the real-space density fields reconstructed from them. For the 
simulations, the reconstructed density fields can also be compared
with the real density fields which are plotted in 
Fig.~\ref{fig:CRSmoothOnSG10MpcFig}.
Overall the level of agreement is impressive. Systematic
differences  are visible in certain areas and presumably reflect
noise effects together with systematics arising because
the smoothed \IRAS 1.2~Jy density field which constrained our
initial conditions does not perfectly trace the true mass density field.
Differences may also arise because the
observed peculiar velocities are generated in part by the mass
distribution outside the region where we constrained our initial 
conditions.

Similar mock catalogues provided the basis for a revised likelihood
analysis of peculiar velocities by \citet{Si01}. The 
proper incorporation of nonlinear effects revealed a systematic overestimate
of $\Omega_{\rm m}$ in earlier linear analyses. It then allowed the 
development of unbiased nonlinear methods, which brought the best estimates 
from the \MarkIII and \SFI data to $\Omega_{\rm m}=0.35\pm0.1$.

We publically release the mock catalogues for \MarkIIIDot They are available
upon request from AE or AD (eldar@phys.huji.ac.il). We can also
provide similar mock catalogues tailor-made for other peculiar-velocity 
catalogues upon request.

%%%%%%%%%%%%%%%%%%%%%%%%%%%%%%%%%%%%%%%%%%%%%%%%%%%%%%%%%%%%%%%%%%%%%%%%%%%%%%%%%%%%%%%%%%%%%

%%%%%%%%%%%%%%%%%%%%%%%%%%%%%%%%%%%%%%%%%%%%%%%%%%%%%%%%%%%%%%%%%%%%%%%%%%%%%%%%%%%%%%%%%%%%%

\section[]{Summary and Conclusions}
\label{sec:summary}
 
The goal of the current study has been to carry out physically
consistent simulations of the formation and evolution of the
local galaxy population within the currently dominant CDM 
structure formation paradigm. We have attempted to reproduce
not only the statistics of galaxy properties (luminosity functions,
colour and morphology distributions, Tully--Fisher relations...) and
of galaxy clustering (correlation functions, cluster luminosities...)
but also the actual spatial distribution of clusters and superclusters
within 80 \hMpc of the Milky Way. This is an ambitious undertaking and 
clearly we have been more successful in some aspects of it than in
others.

On small scales the statistics of the dark matter distributions in our 
\lcdm and \tcdm simulations are a very good match to those expected. On
scales beyond a Mpc or two, discrepancies are apparent which
reflect both the relatively small volume simulated and the constraints
we imposed on the initial large-scale fluctuations. For both
simulations the morphology of the dark matter structure on large
scales is a good match to that of the observed 1.2~Jy \IRAS sample
which we used as a constraint. In addition, the most massive clusters 
agree quite well with observed clusters both in position and in mass.

The distributions of properties of individual galaxies are in fair  
agreement with those observed. The Tully--Fisher relations we find for 
our simulated spiral galaxies are, by construction, in excellent 
agreement with observation. Our distributions of morphology and 
colour also resemble those observed and show the correct 
dependence on galaxy environment. Our luminosity functions, on the 
other hand, show significant discrepancies with those emerging 
from the 2dF and SDSS surveys. This problem is substantially reduced 
from K99, the paper on which our modelling is 
based, because of a number of minor improvements we have introduced,
but it is still worse than found by several other groups carrying out similar 
modelling (e.g. \citealt{Co00,Som01}, S00). We conclude 
that this disagreement probably 
reflects an inadequacy in our galaxy modelling, rather than any 
fundamental problem with the paradigm.

The statistics of galaxy clustering in our \lcdm simulation is in
good agreement with that seen both in the far-infrared selected
\PSCZ survey and in the optically selected \UZC survey, at least
as inferred from the autocorrelation functions of the galaxy
populations. For our \tcdm simulation the agreement with the
\PSCZ survey is still quite good, but the amplitude of optical
galaxy correlations is about 40\% lower than in the \UZC survey.

This latter problem may be related to the fact that the galaxy 
populations of our simulated galaxy clusters are in poor agreement
with observation. Our brightest cluster galaxies are mostly about a
magnitude too bright, while other bright cluster galaxies are about a 
magnitude too faint. Overall our clusters have mass-to-light ratios
at least a factor of 2 larger than observed. This difficulty is
apparently a consequence of the details of our galaxy formation
modelling. \citet{Dia01}, using the simulations of K99,
find a similar problem with the brightest cluster member but find
their other cluster galaxies to be slightly too {\it bright}, while
S00 find cluster luminosity functions
which agree well in shape with observation, but mass-to-light 
ratios which are somewhat high. S00 show the improvement 
in luminosity function shape and the reduction in central galaxy 
luminosity to be a consequence of the greatly improved resolution of 
their simulation. This allows galaxy merging to be followed explicitly 
rather than inserted ``by hand'' using a dynamical friction model.

Point-by-point comparison of the smoothed density fields of mass
and of optical- and FIR-selected galaxies in our two simulations, as
well as comparison with the observed density field of \PSCZ galaxies,
illustrates both the nature of the biases which relate these
various components and the degree to which our techniques have
been successful in reproducing the observed spatial distribution of
star-forming galaxies. Our simulations produce a remarkably tight
relation between the densities of optically selected galaxies and 
of dark matter. Biases are present but are relatively weak, too
weak in the \tcdm case to be consistent with observation, as already
noted above. The fact that star-forming galaxies avoid rich 
groups and clusters results both in more scatter and in a strong bias
at high density in the relation between FIR galaxy density and either 
mass or optical galaxy density. Comparing density fields between
simulations or with the observations shows the strong relations
introduced by our initial condition constraints as well as biases 
caused by differing amounts of dynamical evolution. The excellent
agreement between our simulated FIR galaxy distributions and the
\PSCZ data is best illustrated by the cross-correlation between the
two which is within 20\% of the autocorrelation of the \PSCZ data
on scales above 5 \hMpcDot

Finally our mock catalogues of peculiar velocity data agree well
with the real \MarkIII data. The
reconstructed density fields from the mock catalogues show a 
close correspondance to each other, to the true density fields 
in the simulations, and to the reconstructed density field from the
real data. This is both a reassuring demonstration that our techniques
have achieved their primary goal, and a signpost to the way our
simulations may be used to calibrate quantitative scientific analysis of
the local galaxy distribution. We will present more
detailed applications in future work, and we release a variety
of galaxy, halo and dark matter catalogues in order that others can use them
also. The data are available at the URL:
http://www.mpa-garching.mpg.de/NumCos/CR/index.html

%%%%%%%%%%%%%%%%%%%%%%%%%%%%%%%%%%%%%%%%%%%%%%%%%%%%%%%%%%%%%%%%%%%%%%%%%%%%%%%%%%%%%%%%%%%%%
		
\section*{Acknowledgements}
	
The simulations presented in this paper were carried out on the T3E
supercomputer at the Computing Center of the Max-Planck-Society in
Garching, Germany. This work was supported in part by a grant from
the German-Israel Science Foundation.
 
\bsp

\label{lastpage}

\bibliographystyle{mnras}

\end{document}